\documentclass[longauth]{aa}
\usepackage{graphicx}
\usepackage{caption}
\usepackage{subcaption}
\usepackage{rotating}
\usepackage{booktabs}
\usepackage{multirow}
\usepackage{xcolor}
\usepackage{txfonts}
\usepackage{natbib}
\bibpunct{(}{)}{;}{a}{}{,}{}
\usepackage[colorlinks=true,allcolors=blue]{hyperref}
\usepackage{url}

\usepackage{amstext}

\begin{document}

   \title{The stratification of ISM properties in the edge-on galaxy NGC~891 revealed by NIKA2 } 
   
   \subtitle{}

  \author{S.~Katsioli\inst{\ref{NOA}, \ref{UOA}}\fnmsep\thanks{\email{s.katsioli@noa.gr}}
  \and  E.~M.~Xilouris  \inst{\ref{NOA}}
  \and  C.~Kramer  \inst{\ref{IRAMF}}
  \and  R.~Adam  \inst{\ref{OCA}}
  \and  P.~Ade  \inst{\ref{Cardiff}}
  \and  H.~Ajeddig  \inst{\ref{CEA}}
  \and  P.~Andr\'e  \inst{\ref{CEA}}
  \and  E.~Artis  \inst{\ref{LPSC}, \ref{Garching}}
  \and  H.~Aussel  \inst{\ref{CEA}}
  \and  M.~Baes  \inst{\ref{Ghent}}
  \and  A.~Beelen  \inst{\ref{LAM}}
  \and  A.~Beno\^it  \inst{\ref{Neel}}
  \and  S.~Berta  \inst{\ref{IRAMF}}
  \and  L.~Bing  \inst{\ref{LAM}}
  \and  O.~Bourrion  \inst{\ref{LPSC}}
  \and  M.~Calvo  \inst{\ref{Neel}}
  \and  A.~Catalano  \inst{\ref{LPSC}}
  \and  C.~J.~R.~Clark \inst{\ref{STSI}}
  \and  I.~De~Looze  \inst{\ref{Ghent}, \ref{UCL}}
  \and  M.~De~Petris  \inst{\ref{Roma}}
  \and  F.-X.~D\'esert  \inst{\ref{IPAG}}
  \and  S.~Doyle  \inst{\ref{Cardiff}}
  \and  E.~F.~C.~Driessen  \inst{\ref{IRAMF}}
  \and  G.~Ejlali  \inst{\ref{Iran}}
  \and  M.~Galametz  \inst{\ref{CEA}}
  \and  F.~Galliano  \inst{\ref{CEA}}
  \and  A.~Gomez  \inst{\ref{CAB}}
  \and  J.~Goupy  \inst{\ref{Neel}}
  \and  C.~Hanser  \inst{\ref{LPSC}}
  \and  A.~Hughes  \inst{\ref{Toulouse}}
  \and  A.~P.~Jones  \inst{\ref{IAS}}
  \and  F.~K\'eruzor\'e  \inst{\ref{Argonne}}
  \and  B.~Ladjelate  \inst{\ref{IRAME}}
  \and  G.~Lagache  \inst{\ref{LAM}}
  \and  S.~Leclercq  \inst{\ref{IRAMF}}
  \and  J.-F.~Lestrade  \inst{\ref{LERMA}}
  \and  J.-F.~Mac\'ias-P\'erez  \inst{\ref{LPSC}}
  \and  S.~C.~Madden  \inst{\ref{CEA}}
  \and  A.~Maury  \inst{\ref{CEA}}
  \and  P.~Mauskopf  \inst{\ref{Cardiff},\ref{Arizona}}
  \and  F.~Mayet  \inst{\ref{LPSC}}
  \and  A.~Monfardini  \inst{\ref{Neel}}
  \and  M.~Mu\~noz-Echeverr\'ia  \inst{\ref{LPSC}}
  \and  A.~Nersesian  \inst{\ref{Ghent}, \ref{NOA}}
  \and  L.~Pantoni  \inst{\ref{CEA}, \ref{IAS}}
  \and  D.~Paradis  \inst{\ref{Toulouse}}
  \and  L.~Perotto  \inst{\ref{LPSC}}
  \and  G.~Pisano  \inst{\ref{Roma}}
  \and  N.~Ponthieu  \inst{\ref{IPAG}}
  \and  V.~Rev\'eret  \inst{\ref{CEA}}
  \and  A.~J.~Rigby  \inst{\ref{Leeds}}
  \and  A.~Ritacco  \inst{\ref{ENS}, \ref{INAF}}
  \and  C.~Romero  \inst{\ref{Pennsylvanie}}
  \and  H.~Roussel  \inst{\ref{IAP}}
  \and  F.~Ruppin  \inst{\ref{IP2I}}
  \and  K.~Schuster  \inst{\ref{IRAMF}}
  \and  A.~Sievers  \inst{\ref{IRAME}}
  \and  M.~W.~L.~Smith  \inst{\ref{Cardiff}}
  \and  J.~Tedros \inst{\ref{IRAME}}
  \and  F.~Tabatabaei  \inst{\ref{Iran}}
  \and  C.~Tucker  \inst{\ref{Cardiff}}
  \and  N.~Ysard  \inst{\ref{IAS}}
  \and  R.~Zylka  \inst{\ref{IRAMF}}
}
  
  \institute{
    National Observatory of Athens, Institute for Astronomy, Astrophysics, Space Applications and Remote Sensing, Ioannou Metaxa and Vasileos Pavlou GR-15236, Athens, Greece
    \label{NOA}
    \and
    Department of Astrophysics, Astronomy \& Mechanics, Faculty of Physics, University of Athens, Panepistimiopolis, GR-15784 Zografos, Athens, Greece
    \label{UOA}
    \and
    Institut de Radioastronomie Millim\'etrique (IRAM), 300 rue de la Piscine, 38400 Saint-Martin-d'H{\`e}res, France
    \label{IRAMF}
    \and
    Universit\'e C\^ote d'Azur, Observatoire de la C\^ote d'Azur, CNRS, Laboratoire Lagrange, France 
        \label{OCA}
    \and
    Cardiff Hub for Astrophysics Research \& Technology, School of Physics \& Astronomy, Cardiff University, Queens Buildings, Cardiff CF24 3AA, UK
    \label{Cardiff}
    \and
    Universit\'e Paris-Saclay, Universit\'e Paris Cit\'e, CEA, CNRS, AIM, 91191, Gif-sur-Yvette, France
    \label{CEA}
    \and
    Univ. Grenoble Alpes, CNRS, Grenoble INP, LPSC-IN2P3, 53, avenue des Martyrs, 38000 Grenoble, France
    \label{LPSC}
    \and        
    Max Planck Institute for Extraterrestrial Physics, Giessenbach-strasse 1, 85748 Garching, Germany
    \label{Garching}
    \and
    Sterrenkundig Observatorium Universiteit Gent, Krijgslaan 281 S9, B-9000 Gent, Belgium
    \label{Ghent}
    \and
    Aix Marseille Univ, CNRS, CNES, LAM (Laboratoire d'Astrophysique de Marseille), Marseille, France
    \label{LAM}
    \and
    Institut N\'eel, CNRS, Universit\'e Grenoble Alpes, France
    \label{Neel}
    \and
    Space Telescope Science Institute, 3700 San Martin Drive, Baltimore, MD 21218, USA
    \label{STSI}
    \and
    Department of Physics and Astronomy, University College London, Gower Street, London WC1E 6BT, UK
    \label{UCL}
    \and 
    Dipartimento di Fisica, Sapienza Universit\`a di Roma, Piazzale Aldo Moro 5, I-00185 Roma, Italy
    \label{Roma}
    \and
    Univ. Grenoble Alpes, CNRS, IPAG, 38000 Grenoble, France 
    \label{IPAG}
    \and
    Institute for Research in Fundamental Sciences (IPM), Larak Garden,19395-5531 Tehran, Iran
    \label{Iran}
    \and
    Centro de Astrobiolog\'ia (CSIC-INTA), Torrej\'on de Ardoz, 28850 Madrid, Spain
    \label{CAB}
    \and
    IRAP, Université de Toulouse, CNRS, UPS, IRAP, 9 avenue colonel Roche, BP 44346, 31028 Toulouse Cedex 4, France
    \label{Toulouse}
    \and
    Universit\'e Paris-Saclay, CNRS, Institut d'astrophysique spatiale, 91405, Orsay, France
    \label{IAS}
    \and
    High Energy Physics Division, Argonne National Laboratory, 9700 South Cass Avenue, Lemont, IL 60439, USA
    \label{Argonne}
    \and  
    Instituto de Radioastronom\'ia Milim\'etrica (IRAM), Granada, Spain
    \label{IRAME}
    \and 
    LERMA, Observatoire de Paris, PSL Research University, CNRS, Sorbonne Universit\'e, UPMC, 75014 Paris, France 
    \label{LERMA}
    \and
    School of Earth and Space Exploration and Department of Physics, Arizona State University, Tempe, AZ 85287, USA
    \label{Arizona}
    \and
    School of Physics and Astronomy, University of Leeds, Leeds LS2 9JT, United Kingdom
    \label{Leeds}3
    \and
    Laboratoire de Physique de l’\'Ecole Normale Sup\'erieure, ENS, PSL Research University, CNRS, Sorbonne Universit\'e, Universit\'e de Paris, 75005 Paris, France 
    \label{ENS}
    \and
    INAF-Osservatorio Astronomico di Cagliari, Via della Scienza 5, 09047 Selargius, IT
    \label{INAF}
    \and
    Department of Physics and Astronomy, University of Pennsylvania, 209 South 33rd Street, Philadelphia, PA, 19104, USA
    \label{Pennsylvanie}
    \and 
    Institut d'Astrophysique de Paris, Sorbonne Université, CNRS (UMR7095), 75014 Paris, France
    \label{IAP}
    \and
    University of Lyon, UCB Lyon 1, CNRS/IN2P3, IP2I, 69622 Villeurbanne, France
    \label{IP2I}
  }
   \date{Received; Accepted}

% \abstract{}{}{}{}
% 5 {} token are mandatory
 
\abstract
  % context heading (optional)
  % {} leave it empty if necessary  
   {As the millimeter wavelength range %still 
   remains a largely unexplored spectral region for galaxies, the IMEGIN large program aims to map the millimeter continuum emission of 22 nearby galaxies at 1.15 and 2~mm.
   }
  % aims heading (mandatory)
   {Using the high-resolution maps produced by the NIKA2 camera, we explore the existence of very cold dust and take possible contamination by free-free and synchrotron emission into account. 
   We study the IR-to-radio emission coming from different regions along the galactic plane and at large vertical distances. 
   }
  % methods heading (mandatory)
   {New observations of NGC~891, using the NIKA2 camera on the IRAM~30m telescope, along with a suite of observations at other wavelengths were used to perform a multiwavelength  study of the spectral energy distribution in the interstellar medium in this galaxy. 
   This analysis was performed globally and locally, using the advanced hierarchical Bayesian fitting code, \texttt{HerBIE}, coupled with the \texttt{THEMIS} dust model.
   }
  % results heading (mandatory)
   {Our dust modeling is able to reproduce the near-IR to millimeter emission of NGC~891, with the exception of an excess at a level of 25\% obtained 
   by the NIKA2 observations in the outermost parts of the disk. 
   The radio continuum and thermal dust emission are distributed differently in the disk and galaxy halo. 
   Different dusty environments are also revealed by a multiwavelength investigation of the emission features. Our detailed decomposition at millimeter and centimeter wavelengths shows that emission at 1~mm is purely originated by dust. Radio components become progressively important with increasing wavelengths.
   Finally, we find that emission arising from small dust grains accounts for $\sim9.5$\% of the total dust mass, reaching up to 20\% at large galactic latitudes. Shock waves in the outflows that shatter the dust grains might explain this higher fraction of small grains in the halo.
}
  % conclusions heading (optional), leave it empty if necessary 
   {NIKA2 observations have proven essential for a complete characterization of the interstellar medium in NGC~891.
   They have been critical to separate the dust, free-free, and synchrotron emission in the various emitting regions within the galaxy. 
   }

   %\keywords{galaxies: individual: NGC~891 --
   %             galaxies: spiral, edge-on --
   %             galaxies: dust, ISM, synchrotron, free-free --
   %             galaxies: IR, submillimeter, millimeter, radio
   %            }
   \maketitle
%
%-------------------------------------------------------------------

\section{Introduction} \label{sec:intro}

Observations of galaxies at different wavelengths reveal the different ingredients of which these objects consist and the different physical mechanisms that are taking place within them. The millimeter (mm) to centimeter (cm) part of the spectral energy distribution (SED) of a galaxy is a very important but still vastly unexplored wavelength range in which many physical mechanisms manifest their presence.

Many studies have detected excess
emission in the submillimeter (submm) to mm that was higher than expected from several current dust models, including contamination by radio continuum, molecular lines, and cosmic background (CMB) fluctuations 
\citep[e.g.,][]{2003A&A...407..159G, 2005A&A...434..867G, 2018ARA&A..56..673G, 2009ApJ...706..941Z, 2011A&A...534A.118P, 2012MNRAS.425..763G, 2013A&A...557A..95R, 2016A&A...590A..56H}. In one of the proposed scenarios, this emission originates from the heating of very cold dust grains (T~<~10~K), but other physical mechanisms have also been proposed, such as temperature dependent emissivity and/or magnetic dust grains \citep[see ][for a review]{2018ARA&A..56..673G}. \cite{2020ApJ...900...53C} have concluded that spiral galaxies with a low mass and metallicity are more likely to show submm and/or mm excess. Because high-resolution observations of galaxies at millimeter wavelengths are limited, the interstellar medium (ISM) must be mapped at these wavelengths to detect this excess in different environments and to better constrain its origin.

The radio continuum emission, decomposed into free-free and synchrotron emission, is also present at mm wavelengths. It carries information on the energetics of electrons accelerated by electrostatic or magnetic forces. The synchrotron emission in galaxies is known to be most prominent at radio wavelengths down to 3 cm, and the average slope to range between 0.7 and 0.75 \citep[see, e.g.,][and references therein]{2018A&A...611A..55K}. This emission dominates the halo of galaxies with cosmic-ray electrons which are transported via diffusion along the extraplanar magnetic fields or streaming and winds \citep[e.g.,][]{2019A&A...632A..12S, 2022MNRAS.517.2990T}. 
Free-free emission, on the other hand, is an ideal tracer of ionized hydrogen clouds (H\textsc{ii} regions) in which free electrons scatter off ions that are produced in these regions. The lack of measurements at this wavelength range (especially in spatially resolved observations) has limited our knowledge about the coexistence of these emission mechanisms in different environments within galaxies \citep[see, e.g.,][]{2018A&A...615A..98M,2021A&A...651A..98F}.   

In addition to the well-known thermal dust and free-free and synchrotron emission components, another peculiar component seen in excess is found to be present in some galaxies at this wavelength range and peaks at $\lambda\approx1$~cm (see \cite{2022EPJWC.25700005B, 2022A&A...663A..65Y}, and references therein). This anomalous microwave emission (AME) is detected in both the diffuse medium and in the more compact clouds and can be explained as emission from spinning dust grains \citep{1998ApJ...494L..19D, 2009MNRAS.395.1055A, 2018ARA&A..56..673G}.

The way in which dust is distributed within the galaxies with respect to the stars has been a topic of research for many years. Radiative transfer models have revealed that the way in which the bulk of the dust is distributed in spiral galaxies can be approximated by an exponential disk with one to two times the stellar disk scale length and half of the stellar disk scale height \citep{1999A&A...344..868X, 2016MNRAS.462..331S, 2017A&A...605A..18C, 2000A&A...362..138P, 2020A&A...637A..25N, 2020A&A...643A..90N, 2022MNRAS.515.5698M}. In addition to this component of the diffuse dust, warmer dust that is mostly concentrated in H\textsc{ii} regions has to be taken into account. In radiative transfer models, this is often approximated with a thinner dust disk \citep{2000A&A...362..138P, 2005A&A...437..447D, 2008A&A...490..461B, 2020A&A...637A..25N, 2020A&A...643A..90N}. Furthermore, far-infrared (FIR) observations \citep[e.g.,][]{2011A&A...531L..11B, 2022MNRAS.515.5698M} and also ultraviolet (UV) maps of edge-on galaxies \citep{2014ApJ...785L..18S} reveal large amounts of extraplanar dust at distances of up to a few kiloparsecs above the plane of the galaxies.

The geometry of nearby edge-on galaxies is ideal for studying the properties of the ISM not only in their disks, but also at large distances above the galactic plane \citep[e.g.,][]{1998A&A...331..894X, 2008A&A...490..461B, 2010A&A...518L..39B, 2012MNRAS.419..895D, 2012MNRAS.427.2797D, 2016A&A...592A..71M}. 
It is very important to quantify how dust grains of different sizes are distributed in the different environments within the galaxies. Dust has a strong impact on galactic observables and substantially affects the star formation activity of a galaxy. Therefore, a detailed knowledge of the dust grains properties (e.g., size and temperature distribution and the composition) as well as their evolution is crucial in order to model the ISM in galaxies. The general picture is that smaller grains (with typical sizes smaller than 15~\AA) are distributed around H\textsc{ii} regions, while large grains, which make up the bulk of the dust material, are distributed more evenly in the disk \citep[][and references therein]{2022HabT.........1G}. \cite{1998MNRAS.300.1006D} produced a simple analytical model for assessing the evolution of dust grains within galaxies of various sizes. It was found that large grains $>0.1~\mu$m can travel larger distances within the galaxies (within a period of 10$^9$ yr) and can be recycled through the halo back to the disk, while small grains do not travel far from the location in which they were formed. 

%%%%%%%%%%%%%%%%%%%%%%%%%%% FIGURE 1
    \begin{figure*}[t]
    \centering
    \captionsetup{labelfont=bf}
    \includegraphics[width=0.9\textwidth]{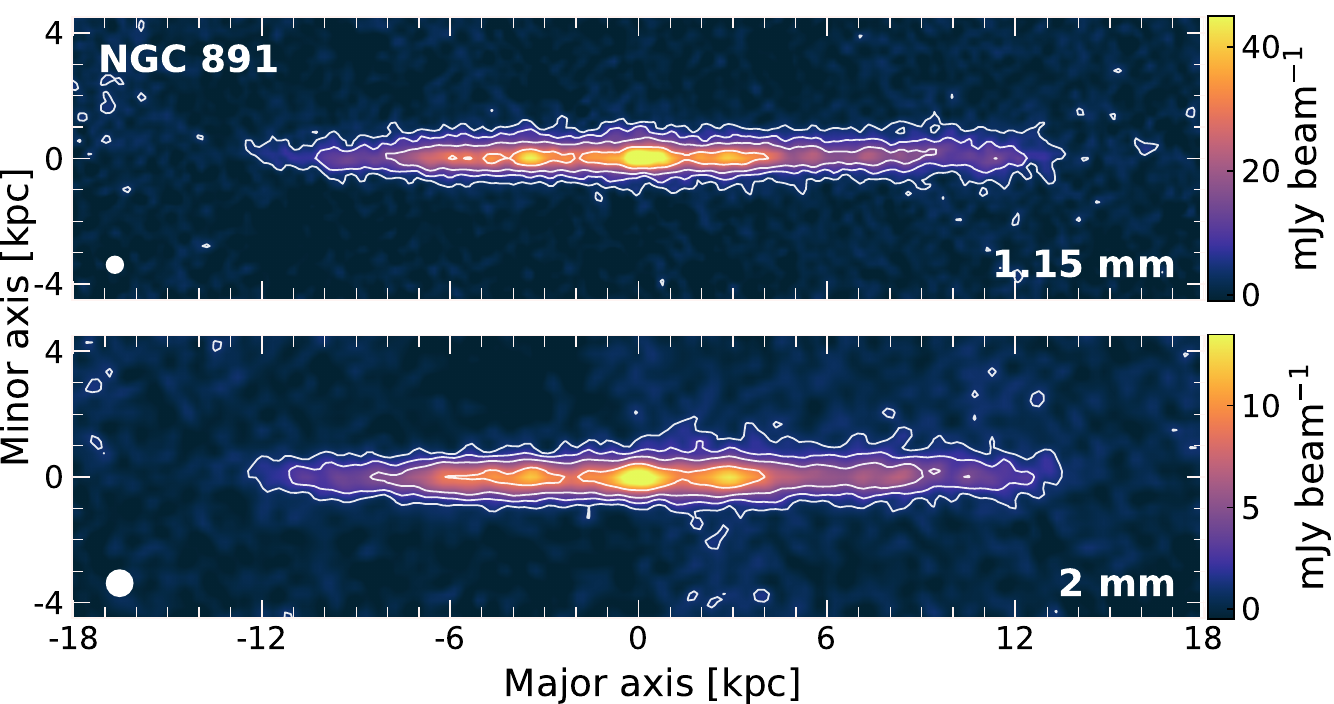}
    \caption{NIKA2 maps 
    of NGC~891 at 1.15~mm (top panel) and 2~mm (bottom panel) with a beam size of 11.1$^{\prime\prime}$ and 17.6$^{\prime\prime}$ ($\sim0.5$~kpc and $\sim0.8$~kpc, respectively; see the white circles in the bottom-left corner in each panel). The maps are centered at RA$_{J2000} = 2^h22^m33^s$, DEC$_{J2000}=+42^{\circ}20^{\prime}53^{\prime\prime}$ and have been rotated by 67.1$^\circ$ counterclockwise (the position angle of the galaxy) for illustration purposes. The surface brightness contours correspond to 3.5, 8, 15, and 30~$\text{times the}$~RMS. The RMS values are 1.0~mJy~beam$^{-1}$ and 0.3~mJy~beam$^{-1}$ at 1.15~mm and 2~mm, respectively. 
    }
    \label{fig:obs}
    \end{figure*}

At a distance of 9.6~Mpc \citep{2011A&A...531L..11B}, NGC 891 is the closest edge-on spiral galaxy. It has been extensively observed over a wide range of wavelengths and was interpreted  with a variety of models. 
It was observed in the X-rays using both \textit{Chandra} and \textit{XMM-Newton} observatories \citep{2013ApJ...762...12H, 2018ApJ...866..126H}. These observations revealed a hot-gas halo around the galaxy with a lower metallicity ($Z\sim0.1$~Z$_\odot$) than the disk metallicity. The disk metallicity is thought to be similar to that of our own Galaxy \citep{2000A&AS..145...83A}. This suggests accretion from the intergalactic medium as the origin of the hot halo. FUV and NUV observations of NGC~891 reveal extended emission above the galactic plane with scale heights of $\approx1.2-2.0$~kpc \citep{2014ApJ...785L..18S}. These observations were interpreted as dust-scattered starlight, indicating the existence of dust grain material at large distances above the plane. Many studies support the scenario that dust is being transported at large distances above the plane through a network of dust chimneys \citep{1997AJ....114.2463H, 2000A&AS..145...83A}, some of which are associated with ionized gas structures.
Deep optical observations have revealed an extended optical halo in NGC~891 that extends up to a few kiloparsecs above the galactic plane \citep{1985ApJ...288..252B}, as well as very low surface brightness features extending up to $\sim40$~kpc from the main body of the disk \citep{2010ApJ...714L..12M}.

\cite{1993A&A...279L..37G} presented the first map of NGC~891 at mm wavelengths (1.3~mm) using the MPIfR 7-channel bolometer array at the \textit{Institut de Radio Astronomie Millim\'etrique} (IRAM) 30m telescope at $12^{\prime\prime}$ resolution with an RMS of 4~-~6~mJy/beam with observations along the major axis out to $\pm10.3$~kpc. These observations were conducted by scanning in azimuth while switching the subreflector/wobbler in azimuth. The resulting map shows a strong correlation of emission between the cold dust material that emits at 1.3~mm and the emission of molecular gas traced by the  CO(2-1) line \citep{1992A&A...266...21G}, but only a weak correlation with H\textsc{i} emission. 
The strong spatial correlation between dust surface mass densities and molecular gas that is traced by the CO(3-2) line was confirmed more recently by \citet{2014A&A...565A...4H}.
They fit modified blackbody (MBB) models to pixel-by-pixel SEDs comprising \textit{Herschel} \citep{2010A&A...518L...1P} and JCMT~-~SCUBA 850~$\mu$m \citep{1999A&A...344L..83I} maps. The derived dust temperatures range between about 17 to 24~K, with an 
average emissivity index of $\beta=1.9$. An extended dust halo was discussed by \citet{2011A&A...531L..11B}, \citet{1998ApJ...507L.125A, 2000A&AS..145...83A}, and \citet{1999A&A...344L..83I}.
\emph{Planck} detected the galaxy at 350, 550, and 850~$\mu$m and at 1.38~mm at resolutions  $\ga5^{\prime}$ \citep{2011A&A...536A...1P}.

In this paper, we profit from the {\it New IRAM Kid Arrays 2} (NIKA2) camera on the IRAM~30m telescope, based on which we map the critical 1.15~mm and 2~mm emission throughout NGC~891 to study the variations in the dust properties within the disk and beyond.
The NIKA2 wavelengths bridge the gap between dust emission at \textit{Herschel} FIR, submm, and the radio emission extending beyond to cm wavelengths.
The observations are part of the {\it Interpreting the Millimetre Emission of Galaxies with IRAM and NIKA2} (IMEGIN) large program (PI: S. Madden), a guaranteed-time large program of 200~hours targeting nearby galaxies (distance smaller than 25~Mpc). 
In Sect.~\ref{sec:Data} we present the NIKA2 observations and data reduction as well as the ancillary data that are used in the subsequent analysis. Sect.~\ref{sec:process} describes the treatment of all the available maps and additional corrections that have to be included so that the maps can be used in the SED modeling analysis, as described in Sect.~\ref{sec:sed}. In the discussion that follows in Sect.~\ref{sec:discussion}, we examine the decomposition of the mm/cm wavelength range into dust thermal emission, free-free emission, and synchrotron emission, the spatial distribution of these components and of the various dust components, as well as the properties and the distributions of the small and large dust grains. Sect.~\ref{sec:conclu} summarizes our conclusions. 

\section{Data} \label{sec:Data}

\subsection{NIKA2 observations and data reduction} \label{sec:obs}

The observations were conducted with the NIKA2 camera at the IRAM~30m telescope at Pico Veleta in the Spanish Sierra Nevada at an altitude of 2850~m. NIKA2 \citep{adam2018,calvo2016,bourrion2016} is a focal-plane camera that observes simultaneously in the two continuum wavebands at 1.15 and 2~mm by means of a dichroic mirror, with angular resolutions of $11.1^{\prime\prime}$ and $17.6^{\prime\prime}$, respectively. There are two 1.15~mm arrays, each made up of 1140 superconducting kinetic inductance detectors (KIDs). The 2~mm array is made of 616 KIDs. They fill the $6.5^{\prime}$ diameter instantaneous field of view (FoV) in each case. The transmission bands of the detectors are broad, with a $\sim50$~GHz full width at half maximum (FWHM) transmission at both wavelengths. The noise-equivalent flux densities (NEFDs) are 33~mJy~s$^{1/2}$ at 1.15~mm and 8~mJy~s$^{1/2}$ at 2~mm. The commissioning campaign was completed in April 2017; the NIKA2 calibration and its performance were presented in \citet[]{2020A&A...637A..71P}. 

%%%%%%%%%%%%%%%%%%%%%%%%%%%%%% TABLE 1
\begin{table*}[t]
\centering
\captionsetup{labelfont=bf, labelformat=simple}
\begin{tabular}{lccccc}
\toprule \toprule
{\bf Telescope (Band)}                                 & {\bf Central } & {\bf Resolution}                  & {\bf Pixel size}                & {\bf Luminosity~$\pmb{\pm}$~RMS }      & \textbf{$\pmb{\pm}$ Calibration } \\
   & {\bf wavelength $\pmb{(\mu}$m)} &                  &                & {\bf (L$\pmb{_{\odot}}$)}      & \textbf{uncertainty}\\ \midrule
WISE (W1) \textsuperscript{(a)}                 & 3.35                            & 6.1$^{\prime\prime}$              & 1.4$^{\prime\prime}$           & (5.02~$\pm$~0.15)~$\times$~10$^9$ \textsuperscript{(b)}    &      3.2\% \textsuperscript{(m)}                    \\
\emph{Spitzer} (IRAC1) \textsuperscript{(a)}           & 3.56                            & 1.7$^{\prime\prime}$              & 0.6$^{\prime\prime}$           & (5.07~$\pm$~0.14)~$\times$~10$^9$ \textsuperscript{(b)}              &      10.2\% \textsuperscript{(m)}        \\
\emph{Spitzer} (IRAC2) \textsuperscript{(a)}           & 4.51                            & 1.7$^{\prime\prime}$              & 0.6$^{\prime\prime}$           & (2.67~$\pm$~0.08)~$\times$~10$^9$ \textsuperscript{(b)}           &         10.2\% \textsuperscript{(m)}         \\
WISE (W2) \textsuperscript{(a)}                 & 4.6                             & 6.4$^{\prime\prime}$              & 1.4$^{\prime\prime}$           & (2.41~$\pm$~0.09)~$\times$~10$^9$ \textsuperscript{(b)}           &        3.5\% \textsuperscript{(m)}           \\
\emph{Spitzer} (IRAC3) \textsuperscript{(a)}           & 5.76                            & 1.9$^{\prime\prime}$              & 0.6$^{\prime\prime}$           & (5.57~$\pm$~0.12)~$\times$~10$^9$ \textsuperscript{(b)}            &        10.2\% \textsuperscript{(m)}         \\
\emph{Spitzer} (IRAC4) \textsuperscript{(a)}           & 8.                              &                                   &                                & (6.86~$\pm$~0.21)~$\times$~10$^9$ \textsuperscript{(b)}          &          10.2\% \textsuperscript{(m)}         \\
WISE (W3) \textsuperscript{(a)}                 & 11.56                           & 6.5$^{\prime\prime}$              & 1.4$^{\prime\prime}$           & (3.51~$\pm$~0.19)~$\times$~10$^9$ \textsuperscript{(b)}           &       5\% \textsuperscript{(m)}           \\
WISE (W4) \textsuperscript{(a)}                 & 22.09                           & 12$^{\prime\prime}$               & 1.4$^{\prime\prime}$           & (2.41~$\pm$~1.47)~$\times$~10$^9$ \textsuperscript{(b)}           &       7\% \textsuperscript{(m)}          \\
\emph{Spitzer} (MIPS1) \textsuperscript{(a)}           & 24.                             & 6$^{\prime\prime}$                & 1.5$^{\prime\prime}$           & (2.07~$\pm$~0.03)~$\times$~10$^9$ \textsuperscript{(b)}            &       4\% \textsuperscript{(m)}          \\
\emph{Spitzer} (MIPS2)                    & 70.                             &                                   &                                & (1.21~$\pm$~0.12)~$\times$~10$^{10}$ \textsuperscript{(b)}         &        5\% \textsuperscript{(m)}        \\
\emph{Herschel} (PACS-Blue) \textsuperscript{(a)}          & 70.                             & 6$^{\prime\prime}$                & 2$^{\prime\prime}$             & (1.39~$\pm$~0.13)~$\times$~10$^{10}$ \textsuperscript{(b)}       &      5.4\% \textsuperscript{(m)}             \\
\emph{Herschel} (PACS-Green) \textsuperscript{(a)}          & 100.                            & 8$^{\prime\prime}$                & 3$^{\prime\prime}$             & (2.10~$\pm$~0.13)~$\times$~10$^{10}$ \textsuperscript{(b)}      &       5.4\% \textsuperscript{(m)}             \\
\emph{Herschel} (PACS-Red) \textsuperscript{(a)}          & 160.                            & 12$^{\prime\prime}$               & 4$^{\prime\prime}$             & (1.63~$\pm$~0.09)~$\times$~10$^{10}$ \textsuperscript{(b)}        &       5.4\% \textsuperscript{(m)}           \\
\emph{Herschel} (SPIRE-PSW) \textsuperscript{(a)}         & 250.                            & 18$^{\prime\prime}$               & 6$^{\prime\prime}$             & (5.14~$\pm$~0.42)~$\times$~10$^9$ \textsuperscript{(b)}           &        5.9\% \textsuperscript{(m)}          \\
\emph{Planck} (HFI1)                    & 350.                            &                                   &                                & (1.65~$\pm$~0.11)~$\times$~10$^9$ \textsuperscript{(f)}           &       4.3\% \textsuperscript{(m)}          \\
\emph{Herschel} (SPIRE-PMW) \textsuperscript{(a)}         & 350.                            & 25$^{\prime\prime}$               & 8$^{\prime\prime}$             & (1.56~$\pm$~0.20)~$\times$~10$^9$ \textsuperscript{(b)}          &      5.9\% \textsuperscript{(m)}            \\
JCMT (SCUBA-450)        & 450.                            &             &         & (6.11~$\pm$~0.92)~$\times$~10$^8$ \textsuperscript{(e)}        &      10\%  \textsuperscript{(n)}             \\
\emph{Herschel} (SPIRE-PLW)          & 500.                            &                                   &                                & (4.35~$\pm$~0.24)~$\times$~10$^8$ \textsuperscript{(b)}           &      5.9\% \textsuperscript{(m)}           \\
\emph{Planck} (HFI2)                   & 550.                            &                                   &                                & (3.02~$\pm$~0.08)~$\times$~10$^8$ \textsuperscript{(f)}            &   4.2\% \textsuperscript{(m)}             \\
\emph{Planck} (HFI3)                   & 850.                            &                                   &                                & (4.85~$\pm$~0.23)~$\times$~10$^7$ \textsuperscript{(f)}             &  0.9\% \textsuperscript{(m)}             \\
JCMT (SCUBA-850)         & 850.                            &             &         & (4.67~$\pm$~0.71)~$\times$~10$^7$ \textsuperscript{(e)}        &       10\% \textsuperscript{(n)}            \\
{\bf IRAM (NIKA2-1.15mm)} \textsuperscript{(b)} & {\bf 1150.}                     & {\bf 11.1$\pmb{^{\prime\prime}}$} & {\bf 3$\pmb{^{\prime\prime}}$} & {\bf (1.56~$\pmb{\pm}$~0.03)~$\pmb{\times}$~10$^{\pmb{7}}$} \textsuperscript{(b)}  &    {\bf 8\%} \textsuperscript{(o)}   \\
\emph{Planck} (HFI4)                    & 1380.                           &                                   &                                & (8.68~$\pm$~0.84)~$\times$~10$^6$ \textsuperscript{(f)}            &       0.5\% \textsuperscript{(m)}          \\
{\bf IRAM (NIKA2-2mm)} \textsuperscript{(b)}    & {\bf 2000.}                     & {\bf 17.6$\pmb{^{\prime\prime}}$} & {\bf 4$\pmb{^{\prime\prime}}$} & {\bf (1.15~$\pmb{\pm}$~0.19)~$\pmb{\times}$~10$^{\pmb{6}}$} \textsuperscript{(b)}  &     {\bf 6\%} \textsuperscript{(o)}   \\
AMI \textsuperscript{(c)}                       & 19341.                          & 24$^{\prime\prime}$               & 5$^{\prime\prime}$             & (5.33~$\pm$~0.52)~$\times$~10$^4$ \textsuperscript{(b)}              &     10\% \textsuperscript{(c)}        \\
OVRO                                            & 28018.                          &                                   &                                & (4.76~$\pm$~0.31)~$\times$~10$^4$ \textsuperscript{(f)}         &       10\% \textsuperscript{(n)}          \\
\emph{Effelsberg~100m}                                           & 28018.                          &                                   &                                & (4.66~$\pm$~0.80)~$\times$~10$^4$ \textsuperscript{(g)}         &      10\% \textsuperscript{(n)}            \\
\emph{Effelsberg~100m}                                           & 28416.                          &                                   &                                & (5.54~$\pm$~0.30)~$\times$~10$^4$ \textsuperscript{(h)}           &        10\% \textsuperscript{(n)}        \\
\emph{Effelsberg~100m}                                           & 34459.                          &                                   &                                & (4.27~$\pm$~0.57)~$\times$~10$^4$ \textsuperscript{(i)}           &      10\% \textsuperscript{(n)}          \\
EVLA (C-band) \textsuperscript{(d)}             & 49965.                          & 15.4$^{\prime\prime}$             & 2$^{\prime\prime}$             & (4.30~$\pm$~0.52)~$\times$~10$^4$ \textsuperscript{(b)}                  &    2\% \textsuperscript{(d)}     \\
WSRT                                            & 60019.                          &                                   &                                & (4.15~$\pm$~0.43)~$\times$~10$^4$ \textsuperscript{(j)}                 &     10\% \textsuperscript{(n)}     \\
GBT                                             & 61813.                          &                                   &                                & (3.48~$\pm$~0.42)~$\times$~10$^4$ \textsuperscript{(k)}                &     10\% \textsuperscript{(n)}     \\
\emph{Effelsberg~100m}                                           & 62457.                          &                                   &                                & (3.99~$\pm$~0.41)~$\times$~10$^4$ \textsuperscript{(l)}           &       10\% \textsuperscript{(n)}        \\ 
\emph{Effelsberg~100m}                                           & 63114.                          &                                   &                                & (4.09~$\pm$~0.41)~$\times$~10$^4$ \textsuperscript{(g)}            &      10\% \textsuperscript{(n)}       \\
\bottomrule
\end{tabular}
\caption{NGC~891 observing and map parameters and the total luminosities. The NIKA2 observations presented here for the first time are highlighted in bold. The entries with available resolution and pixel size information are the maps we used for the resolved galaxy analysis, while the other observations were used for the global analysis. The indices in parentheses in the first column indicate the repository from which the maps were retrieved, while those in the fifth and sixth columns indicate the references that provide the luminosities and uncertainties: \textsuperscript{(a)}~DustPedia database; \textsuperscript{(b)}~this work; \textsuperscript{(c)}~\cite{2018A&A...615A..98M}; \textsuperscript{(d)}~\cite{2015AJ....150...81W}; \textsuperscript{(e)}~\cite{1998ApJ...507L.125A}; \textsuperscript{(f)}~\cite{1983AJ.....88.1736I}; \textsuperscript{(g)}~\cite{1982A&A...116..164G}; \textsuperscript{(h)}~\cite{1995A&A...302..691D}; \textsuperscript{(i)}~\cite{1979A&A....77...25B}; \textsuperscript{(j)}~\cite{1978A&A....62..397A}; \textsuperscript{(k)}~\cite{1991ApJS...75.1011G};\textsuperscript{(l)}~\cite{2009ApJ...693.1392S}; \textsuperscript{(m)}~\cite{2021A&A...649A..18G}; \textsuperscript{(n)}~assumed;  \textsuperscript{(o)}~\href{https://publicwiki.iram.es/PIIC/}{https://publicwiki.iram.es/PIIC/}.
 }
\label{tab:photometry}
\end{table*}

NGC~891 was observed on 21 October 2019, 11 December 2019, 15 January 2020, and 17 January 2020.  A total of 7~hours of telescope time were used to conduct 12 on-the-fly (OTF) scans on NGC~891, alternating the scanning directions of each map between $\pm45^{\circ}$ relative to the major axis of the galaxy to minimize residual stripping patterns in the maps. The central detectors of each array covered an area of $23.4^{\prime}\times10.4^{\prime}$.  Scanning was performed with a speed of 40$^{\prime\prime}$s$^{-1}$ and a spacing of 20$^{\prime\prime}$ between the subscans. 
The observations were taken under stable weather conditions, with the 225~GHz zenith opacities varying between 0.1 and 0.25 with a mean of 0.16, corresponding to 2.7~mm of precipitable water vapor (pwv). Pointing and focus observations and corrections were carried out on nearby quasars about every one or two hours. 
Observations of primary and secondary calibrators were conducted throughout the pool observing campaign.
Dedicated observations of the sky opacity by scanning in elevations were also conducted, but were not used in the data reduction presented here.

Half of the observations were obtained in the UT range 20:00 and 04:00, indicating that the main beams are likely to be stable at their nominal values measured in the commissioning campaigns \citep[][see their Fig.~12]{2020A&A...637A..71P} with half power beam widths (HPBWs) of 11.1$^{\prime\prime}$ and 17.6$^{\prime\prime}$. The remainder of scans were observed in the afternoon between 15:00 and 19:00~UT when the HPBWs often tend to degrade slightly to $\sim12.5^{\prime\prime}$ and $18^{\prime\prime}$, respectively.

The observations were coadded and calibrated using the {\tt piic/gildas}\footnote{\href{https://publicwiki.iram.es/PIIC/} {\label{piic} https://publicwiki.iram.es/PIIC/}} software in the final version of the calibration database with the data associated files (DAFs) \citep{2013ascl.soft03011Z,berta-zylka2022}. {\tt piic} allows for several free parameters to guide the data reduction. They were varied to find the optimum setting, trying to avoid biases and minimize map artifacts. For NGC~891, we ran {\tt piic} (version of 3/2021) with 40 iterations, using third-order polynomials to fit the subscan timelines, and a threshold for the signal-to-noise ratio of 2 and 4 for 1.15~mm and 2~mm, respectively, above which a pixel was considered as a source and was protected in each iteration. To guide the algorithm, the inner part of the source was masked using an ellipse of $180^{\prime\prime}\times20^{\prime\prime}$ rotated by the position angle of NGC~891 and centered on the nucleus.  
The opacities that were measured every 5 minutes by an on-site taumeter working at 225~GHz were interpolated and used to correct for atmospheric absorption at the NIKA2 wavelengths. The extended emission seen by the 30m error beams at 1.15~mm and 2~mm was filtered out through the process. 

To estimate the total absolute flux uncertainties of the NGC~891 observations, we took an absolute flux uncertainty of the primary planets of 5\% into account \citep[see][and refereces therein]{2020A&A...637A..71P}, added in quadrature, with the observed relative RMS scatter on the calibrators as reduced with {\tt piic}\footnoteref{piic}. 
The average RMS during the three pool weeks of observations during day and night, weighted with the number of scans per week, was 5.5\% at 1.15~mm and 2.1\% at 2~mm. This resulted in a total absolute flux uncertainty of 8\% at 1.15~mm and 6\% at 2~mm (see Table~\ref{tab:photometry}).

The final data were projected onto a grid with pixel sizes of 3$^{\prime\prime}$ and 4$^{\prime\prime}$ for 1.15~mm and  2~mm, respectively, to create the final maps (Fig.~\ref{fig:obs}). Both maps show a similar morphology in which the disk extends about $\pm13$~kpc along the major axis from the center of the galaxy and about $\pm1$~kpc above and below the plane at the $3.5~$RMS level. The disk is hardly resolved in the vertical direction.
The inner disk shows regions of enhanced emission. The most pronounced region lies in the direction of the bulge of the galaxy, and two secondary fainter regions lie at $\sim\pm3$~kpc galacto-centric distance along the major axis. Furthermore, fainter blobs of emission are present at various positions in the disk. The outermost regions of the disk at these wavelengths show indications of a warped morphology. This morphology is also evident in the \ion{H}{I} map at the same spatial scales as the NIKA2 maps, and it becomes more prominent at larger radial distances up to $\sim24$~kpc from the center of the galaxy. In general, the emission at the NIKA2 wavelengths agrees well with other wavelengths at FIR and submm, even in the outermost regions of the galaxy. A more detailed presentation of the morphology of the mm maps and a comparison with other wavelengths follows in Sect.~\ref{sec:discussion}.

The observing strategy (fast scanning at $\pm45^\circ$ relative to the  major axis of NGC~891) and data reduction (low-order polynomials fit to the timelines, etc.) were optimized to retrieve the extended mm emission of NGC~891 in the best possible way. The global SED, constructed with the addition of \emph{Planck} (and \emph{Herschel}) data (discussed in detail in Sect.~\ref{subsec:globalSED}), shows that, indeed, NIKA2 retrieves the full emission (see Fig.~\ref{fig:seds}). This is made possible by the edge-on orientation of a relatively thin disk. The FWHMs perpendicular to the major axis are about $2^{\prime}$ only at the Herschel wavelengths \citep{2011A&A...531L..11B}. On the other hand, the spatial scales at which NIKA2 starts to miss extended emission, which is caused by residual drifts of electronics and atmosphere, are larger than about $4^{\prime}$ \citep{2018A&A...615A.112R, 2020A&A...644A..93K}. A detailed discussion of the NIKA2 transfer function as observed in the nearby face-on galaxy NGC~6946 will be presented in Ejlali et al. (in prep.).

\subsection{Ancillary data} \label{sec:ancilliary}

A set of high-resolution maps of the galaxy is necessary for a resolved analysis. Modern instrumentation at infrared (IR) and submm wavelengths renders the resolved analysis of the dusty ISM  feasible. High-resolution radio maps are needed to constrain the SED of the radio continuum emission locally. For the purposes of the current analysis, we compiled data ranging from the near-infrared (NIR) up to cm wavelengths. We used photometrical data received from space infrared and submm telescopes: the \textit{Spitzer Space Telescope} (SST), the \textit{Wide-field Infrared Survey Explorer} (WISE), \textit{Herschel}, and the \textit{Planck Space Telescope}, as well as ground-based telescopes: the \textit{James Clerk Maxwell Telescope} (JCMT), the \textit{Very Large Array} (VLA), the \textit{Arcminute Microkelvin Imager} (AMI), the \textit{100m Effelsberg Telescope}, the \textit{Owens Valley Radio Observatory} (OVRO), the \textit{Westerbork Synthesis Radio Telescope} (WSRT), and the \textit{Green Bank Telescope} (GBT). A complete list of the telescopes and the detectors is given in Table~\ref{tab:photometry}. 

Enriching the available dataset with the NIKA2 maps, presented in this work (see Sect.~\ref{sec:obs}), allows for a high-resolution study of NGC~891 at approximately kiloparsec scales. In order to perform a multiwavelength study, all the maps were convolved to a common spatial resolution. Considering the need to keep the resolution as high as possible, we considered maps of 25$^{\prime\prime}$ (1.17~kpc) resolution or lower. This excluded the SPIRE~-~500~$\mu$m map of 36$^{\prime\prime}$. Because the submm part of the spectrum is well represented at many wavelengths, omitting the SPIRE~-~500~$\mu$m map from the resolved analysis is not expected to introduce any major uncertainties in constraining the SED.

Infrared and submm maps were collected from the DustPedia database\footnote{\label{dustpedia}\href{http://dustpedia.astro.noa.gr/}{http://dustpedia.astro.noa.gr/}}, and radio maps were compiled from the NASA/IPAC Extragalactic Database (NED)\footnote{\label{ned}\href{https://ned.ipac.caltech.edu/}{https://ned.ipac.caltech.edu/}}. The basic properties of the maps we used are given in Table~\ref{tab:photometry}. In this table, entries with an indication of the resolution and the pixel size refer to the resolved maps that were used in the global and spatially resolved SED analysis of the galaxy. The rest were only used as global measurements in the SED analysis. 
Along with the broadband multiwavelength measurements, we also used the CO(3-2) line emission map with a resolution of $\sim 14^{\prime\prime}$ and a pixel size of 7.3$^{\prime\prime}$ that was obtained with the JCMT \citep{2014A&A...565A...4H}. Last, we used the atomic hydrogen (H\textsc{i}) map at $\sim 20^{\prime\prime}$ resolution and 4$^{\prime\prime}$ pixel size that was observed  with the WSRT telescope \citep{2007AJ....134.1019O}.

\section{Data processing} \label{sec:process}

We corrected all maps for background emission using the Python Toolkit for \texttt{SKIRT} \citep[\texttt{PTS};][]{2020A&A...637A..24V} framework. First, we masked all pixels belonging to NGC~891 within an elliptical aperture. Then, we used the mesh-based background estimation from \texttt{photutils} \citep{Photutils} to estimate the large-scale pixel variations around the galaxy. In this method, a particular image is divided into a grid of rectangular regions in which the background is estimated. Following the method of \cite{2020A&A...637A..24V}, we defined square boxes with a side of six times the FWHM of the image. The background emission and its standard deviation were calculated by interpolating the \texttt{photutils} maps within the central ellipse of the galaxy. Then, the background emission was subtracted from the original image. For the NIKA2 maps, only a small offset was fit and substracted that only accounts for less than 0.6\% of the total emission at 1.15~mm and for less than 0.1\% at 2~mm. After the background was subtracted, the error maps were generated for each waveband. For the error maps, we calculated the pixel-to-pixel noise by measuring the deviation of the pixel values from the smooth background. Maps of the galaxy with a signal-to-noise ratio higher than three in each pixel were then created in order to be used in the subsequent SED modeling analysis. Pixels with fewer than two measurements between 60 and 500~$\mu$m were masked out in order to ensure that there are sufficient data to constrain the dust component.

The observed NIKA2 data may be contaminated by line emission. We corrected the 1.15~mm NIKA2 emission for the strongest possible contaminant, the CO(2-1) line, following the method described in \cite{2012MNRAS.426...23D} using the observed CO(3-2) map. To do this, we assumed a CO(3-2)/CO(2-1) line ratio of 0.43 making use of the CO(3-2)/CO(1-0) ratio of about 0.3 found by \cite{2009ApJ...693.1736W} in the diffuse ISM of other nearby galaxies and the CO(2-1)/CO(1-0) ratio of 0.7 found in molecular clouds \citep[e.g.,][]{2018MNRAS.475.1508P}. The NIKA2 transmission curves are given in \citet[][]{2020A&A...637A..71P}. The CO(2-1) line accounts for 1.8\% of the total flux for the whole galaxy, while on local scales, the contamination varies. The highest values are encountered at the center of the galaxy ($2.5-3.5$\%) and at the emission peaks at either side of the center ($1-3$\%). The CO(2-1) contamination in the rest of the galactic disk ranges from 0.1\% to 1.5\%. 

%%%%%%%%%%%%%%%%%%%%%%%%% FIGURE 2
\begin{figure}[h!]
    \centering
    \captionsetup{labelfont=bf}
    \begin{tabular}{@{}c@{}}
     \includegraphics[width=.48\textwidth]{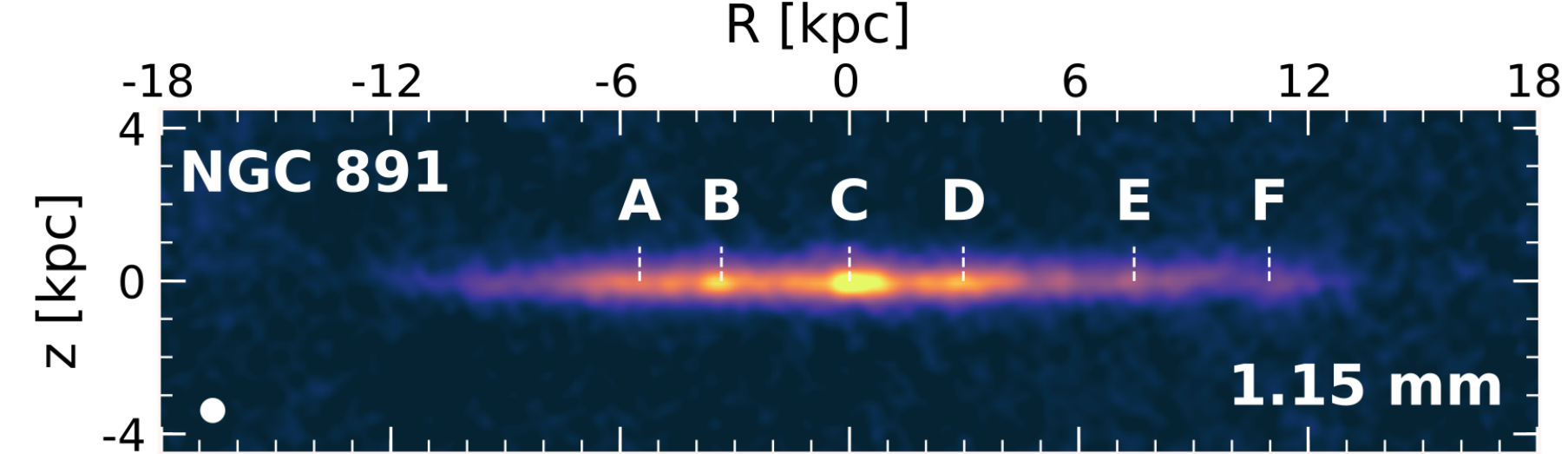} 
     \end{tabular}
    \vspace{0pt}
    \begin{tabular}{@{}c@{}}
        \includegraphics[width=.49\textwidth]{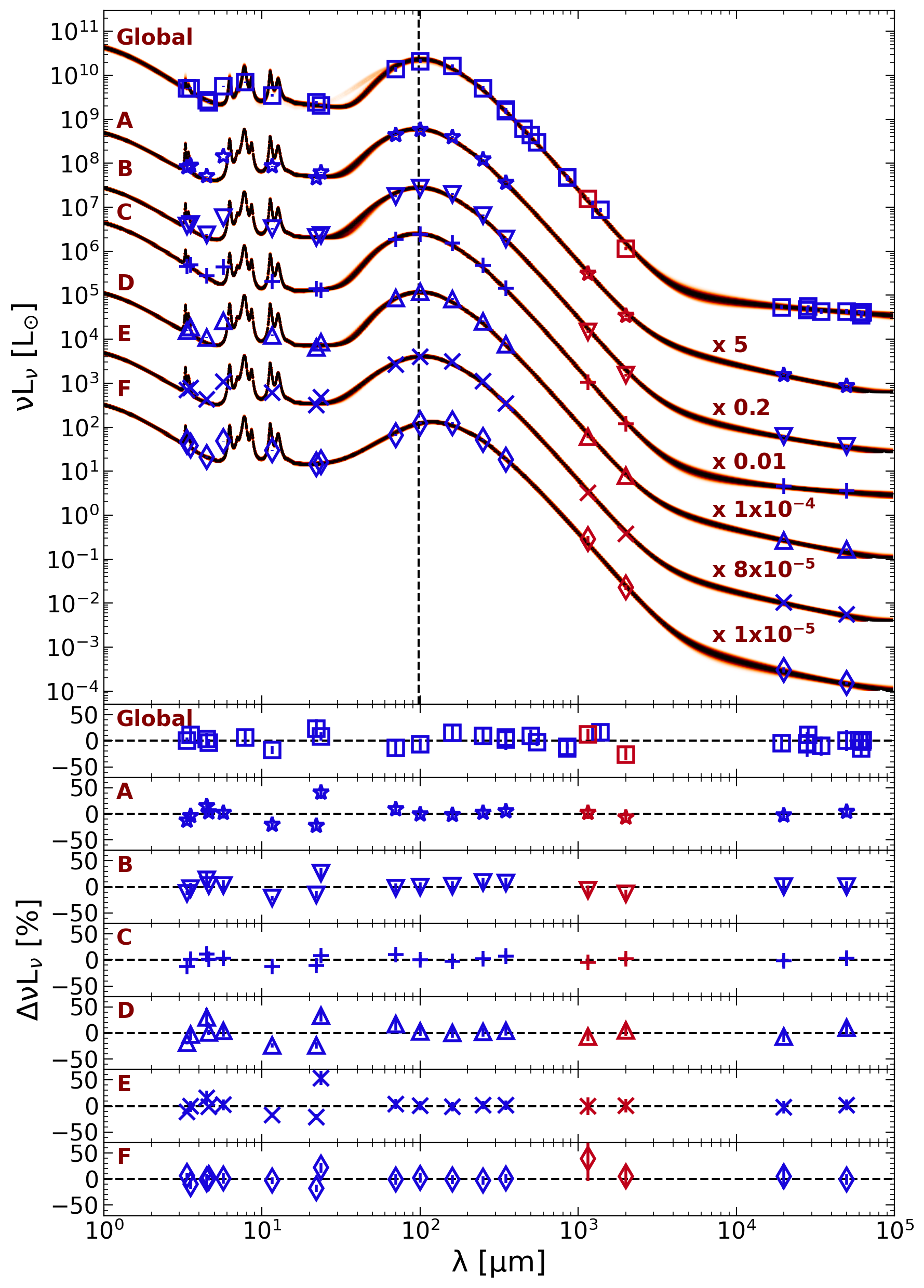} 
    \end{tabular}
    \caption{Spectral energy distributions at different positions throughout the galaxy. The top SED shows the global SED (with \texttt{HerBIE} fitted to the integrated luminosities), and the remaining six SEDs are at the positions A to F that are indicated in the top panel and refer to an area of $0.37\times0.37$~kpc$^2$ each. These positions are centered at $-5.5, -3.4$, 0.0, 3.0, 7.5, and 11~kpc along the major axis of the galaxy (A, B, C, D, E, and F, respectively) and represent the regions of interest discussed in various places in our analysis. The observed luminosities are indicated with the different symbols for each SED. The respective model (and its uncertainty) is presented with the continuous line. NIKA2 luminosities are indicated in red in all the SEDs. The luminosity values are correct only for the global SED, and the rest of the SEDs are scaled by the number indicated next to each model. The vertical dashed line at 98~$\mu$m, the peak wavelength of the IR SED in region A, indicates how the peaks of the rest of the SEDs are placed with respect to this SED. This is an indication of the relative dust temperature difference in the various positions throughout the galaxy.}
    \label{fig:seds}
\end{figure}

Last, because a pixel-by-pixel modeling requires that all the galaxy maps are homogenized in terms of units, resolution, and pixel size, the flux densities were converted into units of monochromatic luminosity (L$_\odot$~Hz$^{-1}$pix$^{-1}$), assuming a distance of 9.6~Mpc \citep{2011A&A...531L..11B} and the corresponding pixel sizes (see Table~\ref{tab:photometry}). Then, we degraded all maps to the 350~$\mu$m map resolution of 25$^{\prime\prime}$ and rebinned them to a common grid with a pixel size of 8$^{\prime\prime}$.

\section{SED modeling} \label{sec:sed}

\subsection{{\tt HerBIE} SED fitting tool} \label{subsec:herbie}

In order to infer the physical parameters of the dust content and also of the radio emission, we used the hierarchical Bayesian inference for dust emission (\texttt{HerBIE}) SED fitting code. \texttt{HerBIE} was described in \cite{2018MNRAS.476.1445G} and \cite{2021A&A...649A..18G}. Using this code, we were able to reveal the integrated galaxy properties, but also to derive the properties of the galaxy on local scales along the line of sight. 

The \texttt{HerBIE} SED fitting code is able to extract information about the basic properties of galaxies using a hierarchical Bayesian approach. This means that the prior distributions are not set before running, but are inferred from the data. This method is more robust than the least-squares approach, for example, in deriving parameters close to the true values and in computing realistic uncertainties. In addition, it is able to eliminate the noise-induced scatter and correlation of the parameters in order to recover the intrinsic scatter and correlation, in contrast to the least-squares method or any other nonhierarchical Bayesian approach. The code samples the probability distribution of the physical parameters by applying a prior distribution controlled by hyperparameters. The probability density function (PDF) of the parameters, which are poorly constrained by the observations, is largely determined by the prior. In contrast to nonhierarchical Bayesian models, the prior we used is itself constrained by the whole distribution of parameters. Poorly constrained parameters are thus inferred from the average distribution of the ensemble of pixels. In total, ten freely varied parameters were considered in our analysis.

%%%%%%%%%%%%%%%%%%%%%% TABLE 2
\begin{table}[t]
\centering
\captionsetup{labelfont=bf, labelformat=simple} %justification=justified
\begin{tabular}{ll}
\toprule \toprule
 \textbf{Parameters}   & \textbf{Global values}   \\ \midrule
    M$_{dust}$~[M$_{\odot}$]        & (3.48~$\pm$~0.22)~$\times$~10$^7$  \\
    M$_{small~grains}$   &  (3.32~$\pm$~0.37)~$\times$~10$^6$ \\
    M$_{large~grains}$   &  (3.15~$\pm$~0.24)~$\times$~10$^7$ \\
    M$_{gas}$/M$_{dust}$   & 261~$\pm$~26        \\
    T$_{dust}$      &   (24.03~$\pm$~0.34)~K    \\
    L$_{star}$~[L$_{\odot}$]       & (6.80~$\pm$~0.37)~$\times$~10$^{10}$   \\
    L$_{dust}$~[L$_{\odot}$]       & (3.39~$\pm$~0.13)~$\times$~10$^{10}$   \\
    <U>~[$2.2\times10^5$~W~m$^{-2}$]    &   2.63~$\pm$~0.95    \\
    q$_{AF}$          & 0.10~$\pm$~0.01           \\
    f$_{ion}$          & 0.21~$\pm$~0.11          \\
    $\beta$    & 1.88~$\pm$~0.47           \\ 
    $\alpha_s$      & 0.79~$\pm$~0.07           \\ 
\bottomrule
\end{tabular}
\caption{Global parameters of NGC~891 as inferred by the \texttt{HerBIE} SED fitting code (see Sect.~\ref{sec:sed}). In adapting the \texttt{THEMIS} model to \texttt{HerBIE}, q$_{AF}$, the mass fraction of a-C(:H) lower than 15~\AA, is an analog to the q$_{PAH}$ in the model of \cite{2007ApJ...657..810D}, and f$_{ion}$, the mass of the a-C(:H) lower than 7~$\AA$ divided by the mass of a-C(:H) lower than 15~\AA, is an analog to the fraction of ionized PAHs. \texttt{HerBIE} incorporates the ISRF of \cite{1983A&A...128..212M} with a mean intensity <U>, normalized in the solar neighborhood, and it parameterizes the Rayleigh-Jeans spectral index $\beta$ and the synchrotron spectral index $\alpha_s$.}
\label{tab:global_sed}
\end{table}

%%%%%%%%%%%%%%%%%%%%%%%%%%%% FIGURE 3
\begin{figure*}[t]
    \centering
    \captionsetup{labelfont=bf}
    \includegraphics[width=0.90\textwidth]{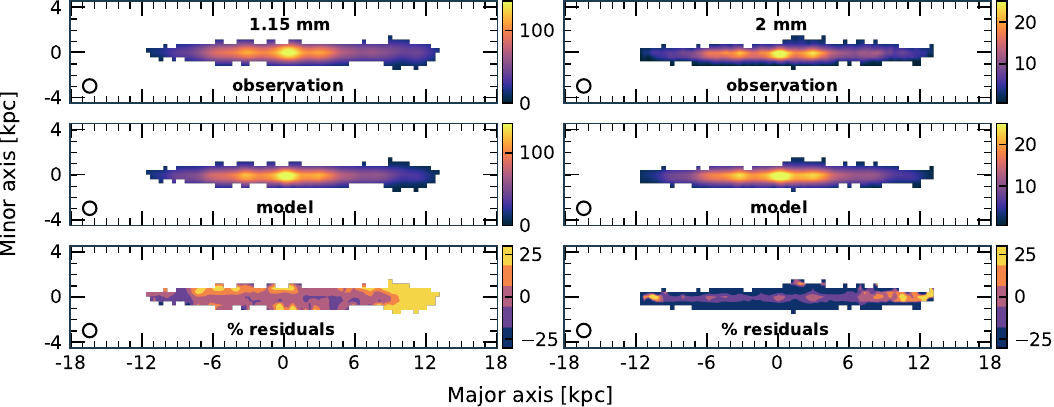}
    \caption{Observed and modeled maps (top and middle panels, respectively) at 1.15 and 2~mm (left and right panels, respectively). The maps are displayed in linear brightness scale. The brightness levels are indicated in the color bars on the right in each panel in units of  mJy per 25$^{\prime\prime}$ beam. The bottom panels indicate the percentage residuals between observation and model at both wavelengths. All maps are at a common resolution of 25$^{\prime\prime}$ , as indicated by the circle in the bottom left corner in each panel. }
    \label{fig:mm_residuals}
\end{figure*}

\subsection{{\tt THEMIS} dust model} \label{subsec:themis}

\texttt{HerBIE} incorporates the advanced dust model \texttt{THEMIS} \citep{2013A&A...558A..62J, 2017A&A...602A..46J} , %dust model 
which consists of core-mantle carbon and silicate grains. The mantles on all grains are photoprocessed, H-poor, atomic-rich carbon. The size-dependent optical properties of this model were constrained as much as possible by laboratory data. %takes into account the presence of both silicate and carbonaceous types of dust grains with size distribution and realistic optical properties \citep{2013A&A...558A..62J, 2017A&A...602A..46J}. 
\texttt{HerBIE} offers the opportunity of modeling different physical conditions by taking into account realistic optical properties, size, starlight intensity distributions, and stochastic heating \citep{1989ApJ...345..230G}. %accounts for the stochastic heating of the small grains and it offers the opportunity to model a combination of different physical conditions. 
The code takes the color correction and the calibration uncertainties of each band into account (see Table~\ref{tab:photometry}). 
For this study, we used the module {\fontfamily{qcr}\selectfont{powerU,}} which includes the formulation for a nonuniformly illuminated dust mixture, the {\fontfamily{qcr}\selectfont{starBB,}} which computes the direct or scattered starlight, and the {\fontfamily{qcr}\selectfont{radio}} module, which combines the free-free and the synchrotron emissions. Hereafter, the parameters of each module are subscribed with these names. Their mathematical formalism is reported extensively in Sect.~2.2 of \cite{2018MNRAS.476.1445G} \citep[see also Sect.~3 in][]{2021A&A...649A..18G}.

\subsection{Global SED} \label{subsec:globalSED}

\texttt{HerBIE} %also 
allows for the use of external parameters as a prior knowledge (e.g., gas mass and metallicity) in the SED fitting. \cite{2018MNRAS.476.1445G} showed that including external parameters in the prior distribution improves the recovery of potential correlations between the external parameters and the dust properties. In the current study, the maps of the atomic and the molecular hydrogen as traced by the H\textsc{i} and CO(3-2) emission lines were included in the prior distribution. 
In addition, after several SED trial fits, it was evident that a reduced abundance of the carbonaceous nanoparticles (smaller than 10~nm) 
by a factor of two had to be applied in order to achieve a better fit at MIPS~-~24~$\mu$m and PACS~-~70~$\mu$m \citep[see Appendix A.2 of][]{2011A&A...536A..88G}.
As a first step, the SED of the whole galaxy was fit using available integrated luminosities from the NIR to radio wavelengths. In this way, the stellar emission, the emission from insterstellar dust, but also the radio emission (taking into account the contributions from both the  free-free and the synchrotron radiation) were constrained on global scales.

The photometry data (luminosity values) for the global SED fitting are listed in Table~\ref{tab:photometry}. The luminosities derived in the current study (labeled ``b'' in Table~\ref{tab:photometry}) were computed within 
an ellipse centered at the center of the galaxy (see Fig.~\ref{fig:obs}), with major and minor axes of 14 and 2.2~kpc, respectively. 
This configuration ensured that the bulk of the emission originating from the disk of the galaxy was measured. The median background level and its RMS value were calculated within an elliptical ring of inner major and minor axes of 18.2 and 7~kpc, and outer major and minor axes of 23 and 12~kpc, respectively.

For the NIKA2 bands, we computed (1.56~$\pm$~0.03)~$\times$~10$^7$~L$_{\odot}$ and (1.15~$\pm$~0.19)~$\times$~10$^6$~L$_{\odot}$ at 1.15~mm and 2~mm, respectively. \cite{1993A&A...279L..37G} used the MPIfR 7-channel bolometer array at the IRAM~30m telescope and measured 5.62~$\times$~10$^6$~L$_{\odot}$ at 1.3~mm, while the \emph{Planck} observation at 1.38~mm measured galactic emission of (8.68~$\pm$~0.84)~$\times$~10$^6$~L$_{\odot}$. The total 1~mm luminosities are difficult to compare because they were not obtained at exactly the same effective frequency and with the same bandwidth. The NIKA2 observations detect more of the faint extended disk because its FoV is larger and the RMS is better by a factor 4~-~6 than in \citet{1993A&A...279L..37G}. With these caveats, the three 1mm measurements agree reasonably well. 
    
The global SED of NGC~891 is presented in Fig.~\ref{fig:seds} (the first SED from top; the luminosity values are plotted as open squares). The width of the model SED is representative of the uncertainty of \texttt{HerBIE} when it derives the SED. 
Overall, the global SED is well constrained by the observations, which is also indicated by the relatively low value of 2.3 of the reduced $\chi^2$ (24 degrees of freedom are considered in this case for a total number of 34 measurements). 
The best-fit parameters derived with the \texttt{HerBIE} code are presented in Table~\ref{tab:global_sed}. The quest for a correct determination of the dust mass in a galaxy is a long-standing problem and is mostly related to the availability of the appropriate measurements. First attempts to measure the dust mass of a galaxy were made in the \textit{Infrared Astronomical Satellite} (IRAS) era, in which observations were only sensitive to the warm dust. This resulted in a strong underestimation of this parameter. For NGC~891, the IRAS-derived dust mass was calculated to be $8.8\times10^6$~M$_{\odot}$ \citep{1990ApJ...359...42D}, scaled to the adopted distance of 9.6~Mpc. The bulk
of the dust could be detected with the advent of higher sensitivity to the cold dust at mm to submm wavelengths through space observatories such as \textit{Infrared Space Observatory} (ISO) and \textit{Herschel}, but also the ground-based telescope, JCMT. This resulted in dust masses that were much higher than previously
thought. Using SCUBA observations, \cite{1998ApJ...507L.125A} computed a dust mass of $5.1\times10^7$M$_{\odot}$ that was later supported by the inclusion of ISO observations at FIR wavelengths \citep[$7.2\times10^7$~M$_{\odot}$;][]{2004A&A...414...45P}. Radiative transfer models that only account for the extinction effects of the stellar light by the dust (and do not take the dust emission into account) have been successful in determining the dust mass quite accurately \citep[$5.7\times10^7$~M$_{\odot}$;][]{1999A&A...344..868X}. 
Recent studies that mostly exploited the \textit{Herschel} capabilities \citep[e.g.,][]{2014A&A...565A...4H, 2021MNRAS.502..969Y} reported dust masses for NGC~891 ranging between $8.5\times10^7$ and $1.1\times10^8$~M$_{\odot}$, depending on the regions considered and the assumptions for the dust emissivity index fit with a MBB, and with dust temperatures ranging between 17~K and 24~K. With our analysis that includes the NIKA2 observations, adopts the \texttt{THEMIS} dust model, and uses the Bayesian approach that \texttt{HerBIE} performed, we derive a dust mass of $3.48\times10^7$~M$_{\odot}$ and a dust temperature of 24~K. 
The derived dust temperature agrees well with previous studies. The dust mass value is lower by a factor of $\sim2-3$  than the values derived with other methods (e.g., MBBs) that only considered the \textit{Herschel} data. This can be explained by the different adopted codes and dust grain models \citep[see, e.g.,][]{2021ApJ...912..103C}.

%%%%%%%%%%%%%%%%%%%%%%%%%%%% FIGURE 4
\begin{figure}[t]
    \centering
    \captionsetup{labelfont=bf}
    \includegraphics[width=0.5\textwidth]{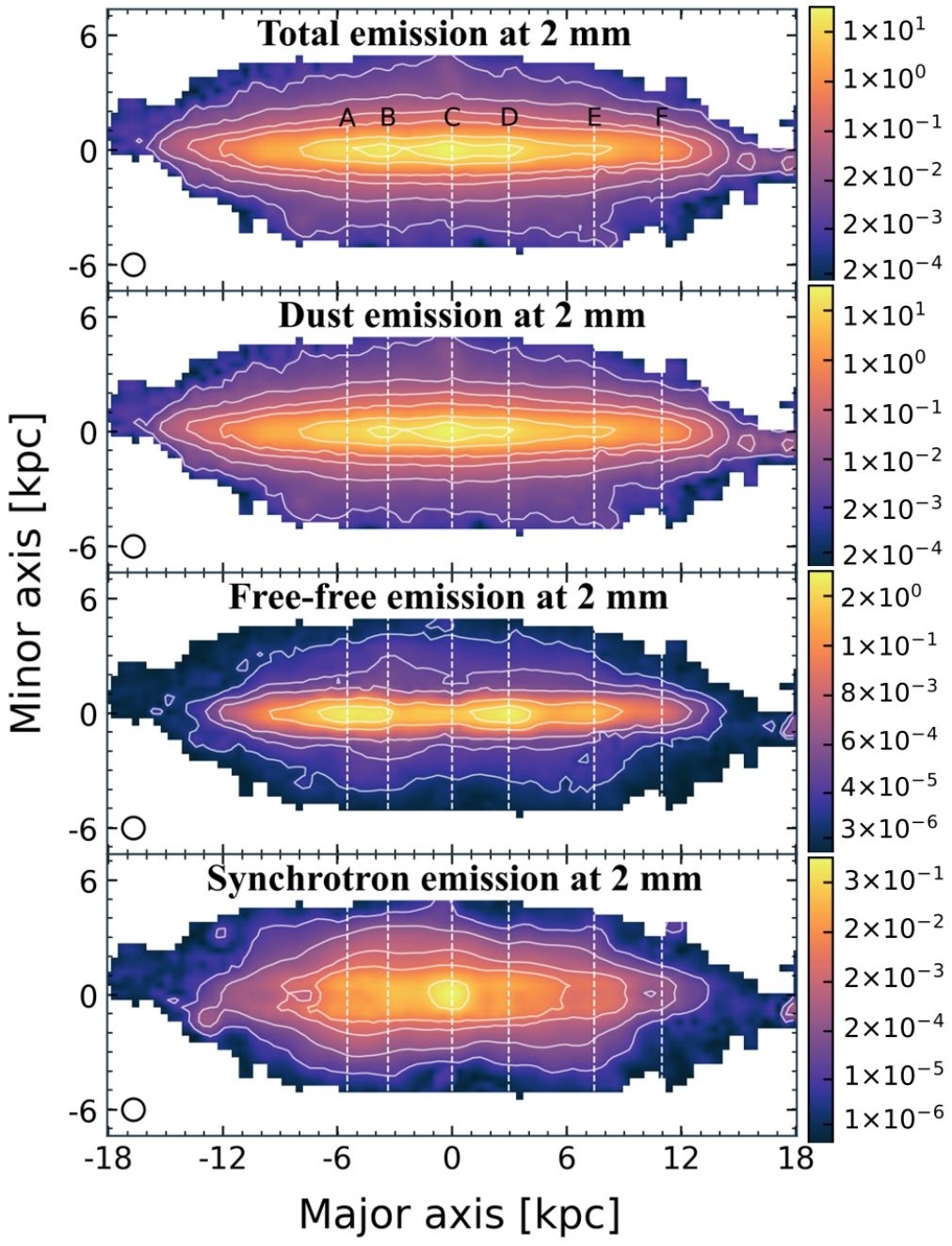}
    \caption{Decomposition of the modeled emission at 2~mm into dust, free-free, and synchrotron emission. The top panel shows the total modeled emission, and the next three panels show the dust, the free-free, and the synchrotron emission maps from top to bottom, respectively. The color bars are in mJy per 25$^{\prime\prime}$ beam. The vertical dotted lines indicate the positions of regions A to F (see Fig.~\ref{fig:seds}).} 
    \label{fig:2mm}
\end{figure}

In order to calculate the total gas mass of NGC~891, we used the WSRT 21~cm line measurements \citep{2007AJ....134.1019O} for the atomic hydrogen mass (M$_{H_\textsc{I}}$) and the JCMT CO(3-2) line measurements \citep{2014A&A...565A...4H}. We adopted a line ratio of CO(3-2)/CO(1-0)~=~0.3 \citep{2009ApJ...693.1736W} to convert the CO(3-2) line measurements into molecular hydrogen mass (M$_{H_2}$). We found $3.05\times10^9$ and $3.81\times10^9$~M$_{\odot}$ for M$_{H_2}$ and M$_{H_\textsc{I}}$ , respectively. Multiplying the sum of the hydrogen mass by a factor of 1.36 to account for the contribution of helium and heavy elements, we find a total gas mass of $(9.08~\pm~0.60)\times10^9$~M$_{\odot}$ and a gas-to-dust mass ratio of 261~$\pm$~26.

\subsection{Spatially resolved SED} \label{subsec:resolvedSED}

In addition to the global SED, we computed the spatially resolved SEDs of individual areas throughout the galaxy. This provided us with information about the variation in the physical parameters in different environments within the galaxy. Examples of the SED models are given in  Fig.~\ref{fig:seds} at six positions in the galaxy (see Fig.~\ref{fig:seds}), and the typical parameters derived with our analysis for these regions are provided in Table~\ref{tab:spatial_sed} (see Appendix~\ref{sec:appendA} for details). The pixels were chosen in such a way as to represent different environments, such as the very center of the galaxy (C), the secondary peaks around the center (B, and D), the positions of enhanced brightness farther out in the disk (A, and E), and a position at the very faint end of the disk (F).
The reduced $\chi^2$ values for the regions A to F are 8.85, 6.72, 3.16, 12.61, 8.59, and 2.53, respectively, considering seven degrees of freedom (17 measurements).
Although the observations compare well with the model, the high residuals found in the outermost regions at mm wavelengths (at 11~kpc from the center; see, e.g., the residuals in region F in Fig.~\ref{fig:seds}) need further investigation. At this position, the data seem to indicate an excess of emission, and the dust appears to be underestimated by \texttt{HerBIE} and \texttt{THEMIS}. In order to better visualize areas with excess emission at mm wavelengths, we compared the observed and modeled maps at 1.15 and 2~mm and computed the respective residual maps, which are shown in Fig.~\ref{fig:mm_residuals}. The residual maps clearly show that although the observed and modeled maps agree well  at the two wavelengths (the residuals are lower than 10\% in most places in the galactic disk), in regions in the outer part of the disk (between 10 and 12 kpc), the observation is higher than the model prediction by more than $\sim 25\%$. This may indicate an additional component of very cold dust in the outer parts of the galaxy. 
Another obvious point from the SEDs in Fig.~\ref{fig:seds} is that the dust gradually becomes colder while moving from the center of the galaxy to the edges. 
The vertical dashed line in the plot, centered at the peak of the dust emission (occurring at 98~$\mu$m) at the center of the galaxy (region C), shows the relative displacement of the peaks of the SEDs of the different regions. It indicates that the SED peaks shift farther away from the center peak at longer wavelengths (colder dust temperatures).

%%%%%%%%%%%%%%%%%%%%%%%%%%%% FIGURE 5
\begin{figure}[t]
    \centering
    \captionsetup{labelfont=bf}
    \begin{tabular}{@{}c@{}}
     \includegraphics[width=.44\textwidth]{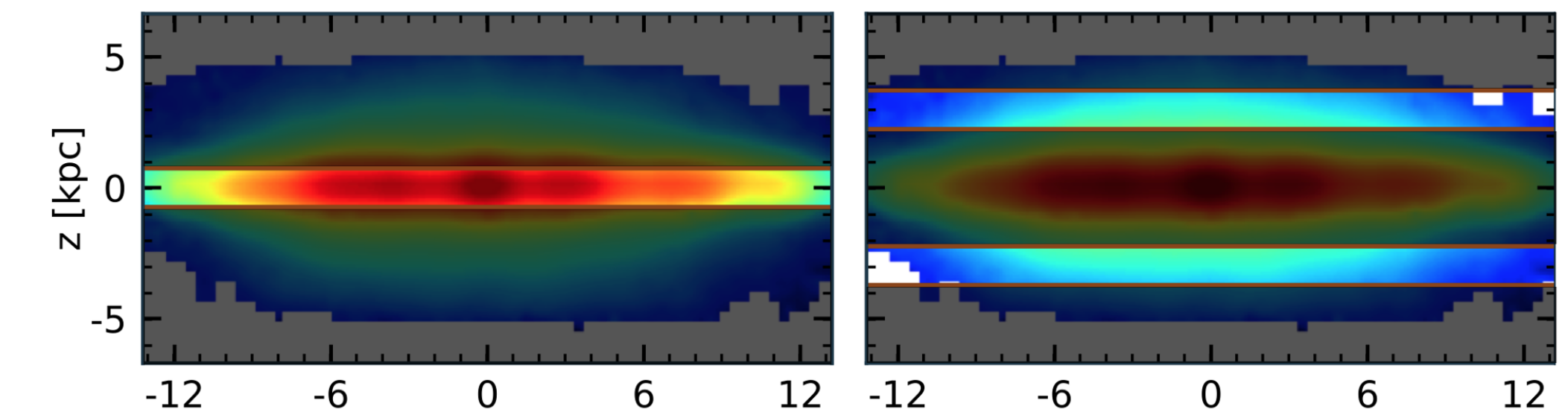} 
     \end{tabular}
    \vspace{0pt}
    \begin{tabular}{@{}c@{}}
        \includegraphics[width=.44\textwidth]{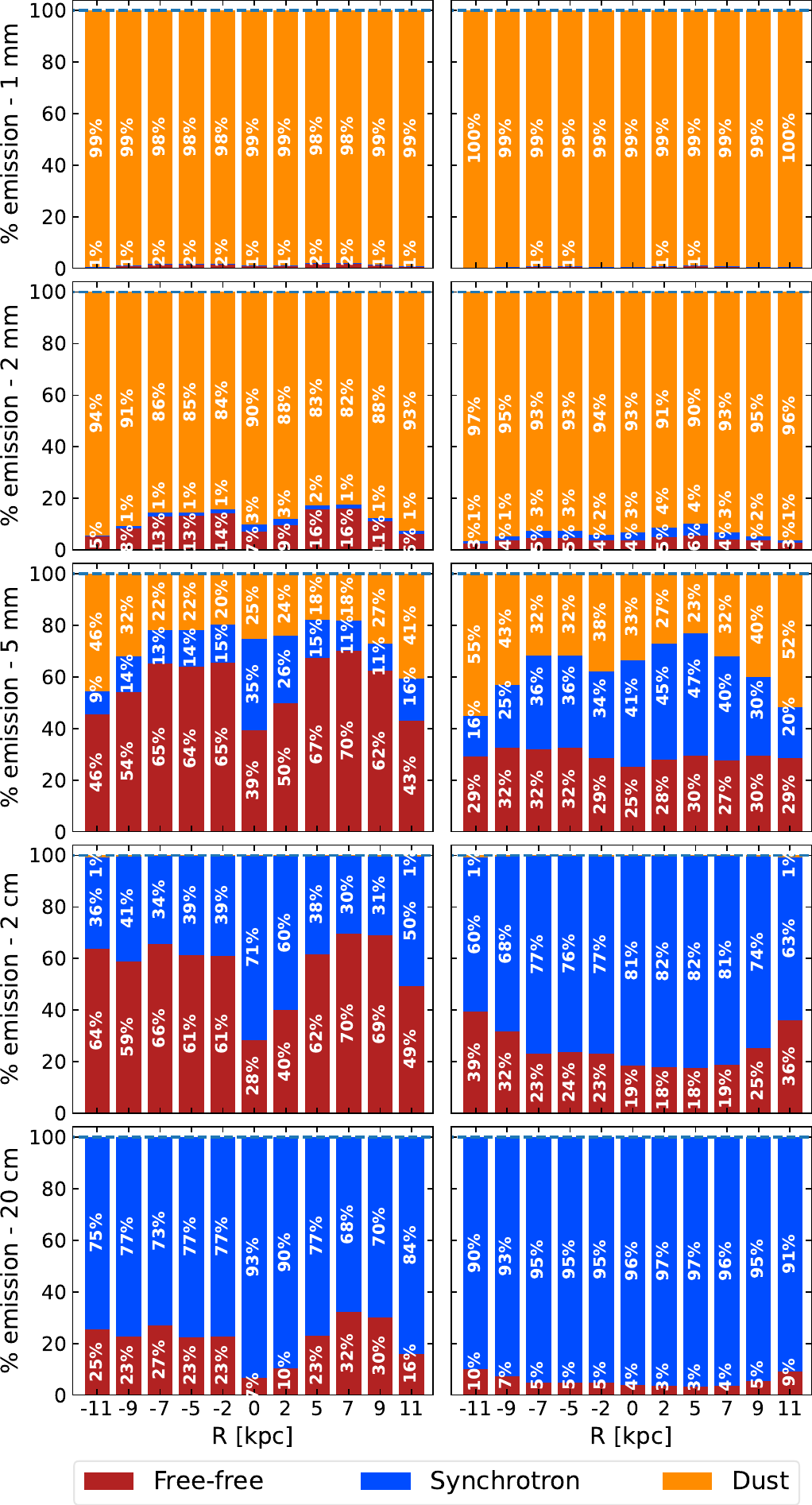} 
    \end{tabular}
    \caption{Emission components that contribute to the total flux at 1, 2, and 5~mm, 2 and 20~cm (top to bottom panels) in the galactic plane (left panels) and in the halo (right panels). The emission percentages for the dust, free-free, and synchrotron emissions are shown as orange, red, and blue bars, respectively. 
    The actual percentage for each emission mechanism at each position is given in numbers in the plots. The top two panels indicate the positions along the major axis of the galaxy where the decomposition was made. For the halo at |z|~$\sim3$~kpc (right panel), the mean values in the two horizontal lanes indicated in the plot were taken into account.}\label{fig:fig8}
\end{figure}

\section{Discussion} \label{sec:discussion}
\subsection{Decomposing the emission in the mm to cm wavelength range} \label{decomp}

The mm to cm wavelength range is a very complex region of the galactic SED in which emission may originate from a variety of mechanisms. The main emission sources are the thermal emission from dust grains (mainly from large dust grains emitting at low temperatures) and the emission of the ionized gas comprising free-free and  synchrotron emission. In addition, other secondary less obvious mechanisms that are difficult to detect may take place, revealing their existence at these wavelengths. These can be very cold dust grains that produce excess emission at mm wavelengths as well as anomalous microwave emission (AME) emitting at cm wavelengths \citep[see][for a review]{2022HabT.........1G}.We have carried out a detailed decomposition of the emitting mechanisms to evaluate their relative importance, both spatially and at different wavelengths.

Fig.~\ref{fig:2mm} presents the modeled emission map at 2~mm. The emission was decomposed using the \texttt{HerBIE} model into dust emission, free-free emission, and synchrotron emission. It is worth noticing here that the modeled map at 2~mm is more extended than the observed emission (Fig.~\ref{fig:obs}). This is because, as described in Sec.~\ref{sec:process}, the produced modeled maps depend on all the available maps that are used in the \texttt{HerBIE} fitting code, which consider the more extended emission that is detected at other wavelengths.

On the global scale of the galaxy, 91\% of the total 2~mm emission arises from cold dust, 5\% from free-free emission, and 4\% from synchrotron emission.  
The bulk of the emission is concentrated along the disk of the galaxy, but a prominent halo component is also present. The dust disk is composed of four prominent components, a central peak C, two secondary peaks B and D, and a diffuse halo component that extends to vertical distances beyond the plane of the disk, to up to 5 kpc. The enhanced dust emission feature above the center of the galaxy (region C) at a distance of $\sim4$~kpc is also interesting. 
\cite{2021MNRAS.502..969Y} also indicated dusty filaments such as this one using image-sharpening techniques (see their Fig.~9), rising up to $\sim4$~kpc above the galactic plane, based on FIR observations. We find that this feature carries 1.95$\times10^5$~M$_{\odot}$ of dust, 14\% of which is small dust grains (see also Sect.~\ref{sec:dust_grains}). 
The dust halo shows a symmetric elliptical distribution that keeps its shape even at large distances away from the disk of the galaxy. 

The free-free emission map shows enhanced emission in the disk, but no obvious emission is detected at the center of the galaxy (region C). In contrast, enhanced free-free emission is seen in regions B and D, but also in regions A and E. The free-free emission at high galactic latitudes seems to follow the general shape of the features in the disk (i.e., a deficit in the central region C and enhancement in regions A, B, D, and E), forming a peanut-shaped halo. The synchrotron emission map, on the other hand, shows the opposite topology with respect to that of the free-free emission, with enhanced emission at the center of the galaxy (region C) that decreases toward regions A, B, D, and E. The synchrotron emission halo maintains the same peanut-shaped distribution as in the case of the free-free emission.    

In Fig.~\ref{fig:fig8} we show the relative contribution of the three main mechanisms at four wavelengths (1~mm, 2~mm, 5~mm, 2~cm, and 20~cm) bracketing the cases where dust thermal emission and radio emission dominate the SED (at 1~mm and 20~cm, respectively). The leftmost panels show the fractions of the different emission mechanisms in the disk of the galaxy (see the central lane indicated in the top map, and the rightmost panels indicate the fractions in the halo of the galaxy). These were calculated as the median values within two parallel lanes above and below the disk of the galaxy centered at 3~kpc from the plane of the disk.

In the disk of the galaxy (leftmost panels in Fig.~\ref{fig:fig8}), the emission at 1~mm clearly mainly arises from thermal dust (at a $98-99$\% level), with negligible contamination from free-free emission ($\sim1-2$\%) and practically no synchrotron emission. At 2~mm, the free-free emission begins to increase and is more prominent at the secondary peaks around the center (regions B and D; with a contribution of $\sim15$\% ), while the synchrotron emission begins to be detected at the center (with a contribution of only $\sim2-3$\% ). At 5~mm, the radio emission becomes the prominent component, and emission from thermal dust is only $\sim20-30$\%, depending on the position inside the galaxy. From the radio emission, the free-free component is the most important with up to $\sim70$\% in the secondary peaks, which drops to $\sim39$\% in the center. The interplay between the two radio emission mechanisms with synchrotron emission filling in the gaps where free-free emission shows a deficit is interesting (e.g., in the center of the galaxy). This becomes more obvious at longer wavelengths (2~cm) where the dust emission is negligible (lower than 1\%). As a result, the center of the galaxy is synchrotron dominated ($\sim$70\%), while the secondary peaks are high in free-free emission ($\sim60$ to $70$\%). At much longer wavelengths (20~cm), synchrotron emission dominates the disk with up to $\sim93$\% in the center and with $\sim75$ to $80$\% in the rest of the disk.      

The galaxy halo (rightmost panels in Fig.~\ref{fig:fig8}) shows significant differences in the way the different emission mechanisms are distributed compared to
the galactic disk. At 1~mm, the difference is negligible because dust emission dominates both regions at $\sim 99$\%. At 2~mm, the halo of the galaxy is still dominated by dust emission at $\sim93$\%, and the free-free and synchrotron emission begin to increase, but at lower level than in the disk (by $\sim4-6$\% and $\sim 2-4$\% for the free-free and the synchrotron emission, respectively). The difference compared to the disk in the distribution along the major axis is also notable. It is flatter in the halo, without obvious central and secondary peaks. At 5~mm, the dust emission in the halo is the weakest component, at a somewhat higher fraction
compared to the disk ($\sim30-40$\%). The free-free emission is at the level of $\sim30$\%, much lower than in the disk, and its distribution is flat, while the synchrotron emission starts to become the dominant emission and contributes up to $\sim47$\%. 
At 2~cm and 20~cm, synchrotron emission has become the dominant emission in the halo of the galaxy and contributes up to $\sim84$\% and $\sim97$\% at 2~cm and 20~cm, respectively. The difference compared to the disk of the galaxy is evident: The contribution of free-free emission is far lower, and the distribution along the major axis is flatter.

%%%%%%%%%%%%%%%%%%%%%%%%%%%% FIGURE 6
\begin{figure}[t]
    \centering
    \captionsetup{labelfont=bf}
    \includegraphics[width=0.5\textwidth]{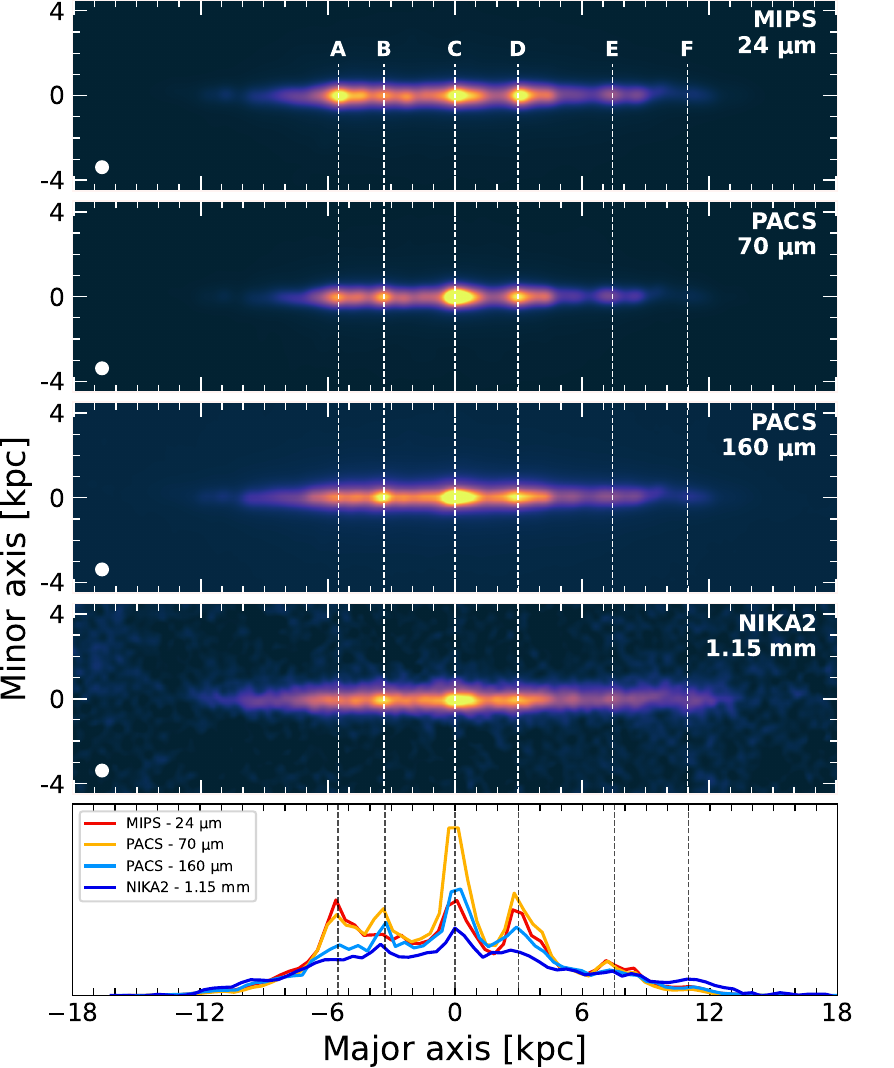}
    \caption{Distribution of different dust components as traced by the MIPS~-~24~$\mu$m, the PACS~-~70~$\mu$m, the PACS~-~160~$\mu$m, and the NIKA2~-~1.15~mm (top to bottom panels, respectively) convolved at an FWHM of 12$^{\prime\prime}$. The profiles along the major axis are plotted in the bottom panel. All profiles are normalized at 6~kpc. The vertical dotted lines indicate the positions of interest introduced in Fig.~\ref{fig:seds}. The colors were scaled in such a way as to detail the disk morphology and not the extended emission above the plane.}    \label{fig:warm_cold}
\end{figure}

%%%%%%%%%%%%%%%%%%%%%%%%%%%% FIGURE 7
\begin{figure}
    \centering
    \captionsetup{labelfont=bf}
    \includegraphics[width=0.49\textwidth]{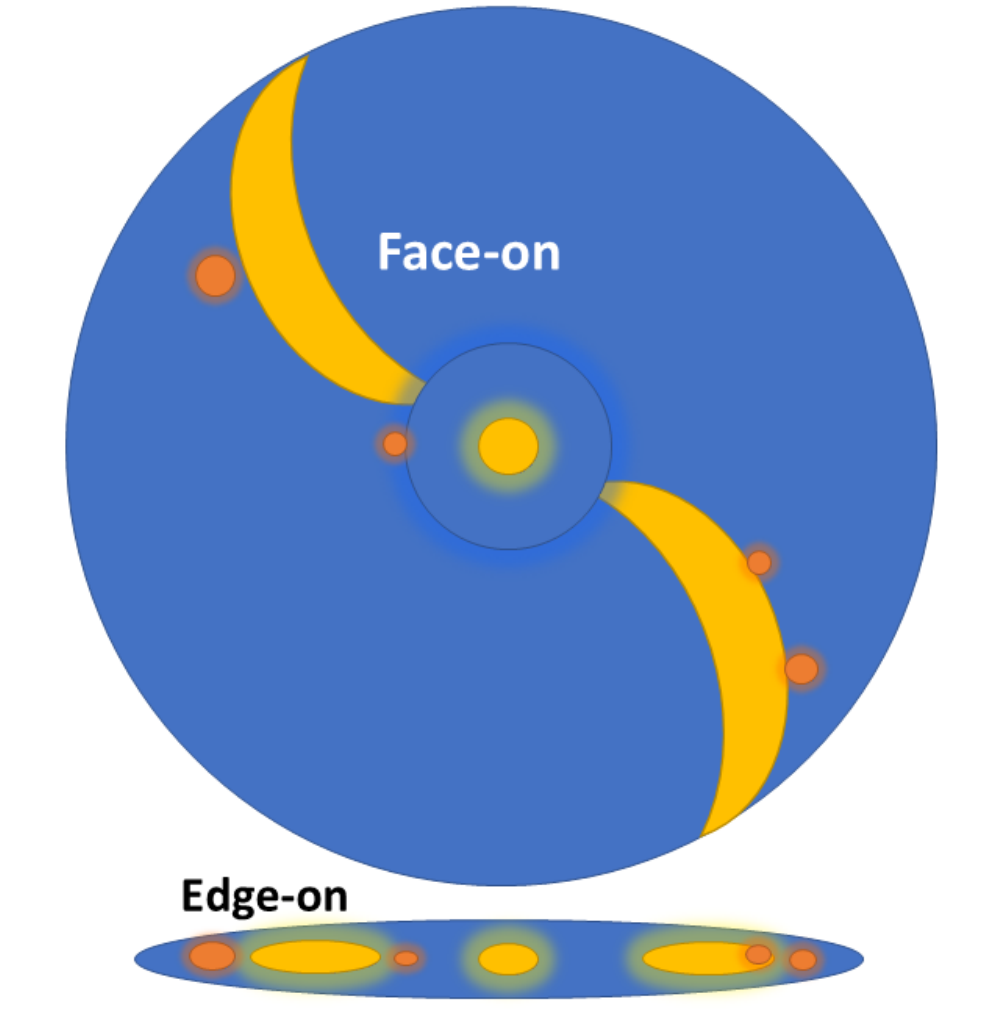}
    \caption{Graphical illustration of the different dust emitting regions in NGC~891 in a face-on and an edge-on configuration (top and bottom, respectively). This illustration presents the diffusely distributed dust along the dust disk (blue), the cold dust along the spiral arms and the bulge (yellow), and warm dust emitting from individual H\textsc{ii} regions (orange).}
    \label{fig:sketch}
\end{figure}

\subsection{Distribution of the warm and cold dust}\label{sec:dust_temp}

In the general picture of the dust distribution in spiral galaxies, the cold dust material is distributed diffusely throughout the disk of the galaxy and along the spiral arms, and the warm dust is mainly found near the H\textsc{ii} regions. This is directly visible in the face-on geometry because disk, spiral arms, and H\textsc{ii} regions can be easily spotted, but they are hard to distinguish in the edge-on configuration. \cite{2010A&A...518L..67K} and \cite{2012A&A...543A..74X} have examined this by studying the distribution of the warm and cold dust in the face-on Local Group galaxy M33, but recent radiative transfer modeling of face-on galaxies also examined the detailed geometry of dust that is diffusely distributed in the disk and also concentrated in compact H\textsc{ii} regions \citep[see, e.g.,][]{2020A&A...637A..24V,2020A&A...637A..25N,2020A&A...638A.150V,2020A&A...643A..90N}.

The edge-on configuration of NGC~891 reveals a dust morphology that is not smoothly distributed, but has obvious enhancements at several positions throughout the disk of the galaxy. This is evident in Fig.~\ref{fig:warm_cold}, in which we plot the MIPS~-~24~$\mu$m, the PACS~-~70~$\mu$m, the PACS~-~160~$\mu$m, and the NIKA2~-~1.15~mm maps. The maps here were convolved to an FWHM of 12$^{\prime\prime}$, the limiting resolution of PACS~-~160~$\mu$m (see Table~\ref{tab:photometry}), and they are meant to reveal the morphology of the central plane of the disk and not the extended halo. The emission along the central spine shows the six substructures introduced in Fig.~\ref{fig:seds}, which are marked with vertical dotted lines in the four maps. The diffuse dust emission is present in all bands, and the different wavelengths show varying relative intensities of the disk substructures.  
Structure C, at the very center of the galaxy, appears in all bands and seems to dominate the bulge region. 
Structures B and D are very prominent at PACS~-~160~$\mu$m and NIKA2~-~1.15~mm and become dimmer at shorter wavelengths (especially B, which is very faint at MIPS~-~24~$\mu$m). On the other hand, structures A and E are bright at MIPS~-~24~$\mu$m and become progressively fainter at increasing 
wavelengths. Because MIPS~-~24~$\mu$m is mostly sensitive to warmer dust, while PACS~-~160~$\mu$m and NIKA2~-~1.15~mm trace the cool  dust, it might be argued that these are three different types of dust environments, with C, the central bulge region, hosting both warm and cold dust grain material, B and D being dominated by cold dust, and A and E mostly composed of warm dust material. 

One scenario that explains the different regions at different wavelengths is presented in Fig.~\ref{fig:sketch} with the schematic of the edge-on view of the galaxy with the predominant dust components and the corresponding face-on orientation of the galaxy. 
Here, the blue color represents the dust that is smoothly distributed in the disk of the galaxy, yellow is the, mostly, cold dust that is distributed along the spiral arms, and orange indicates warm dust in H\textsc{ii} regions. 
This sketch is only meant to describe the dust emission distribution and is kept in a very simplistic form, even though it is known that more complicated structures \citep[e.g., a central bar;][]{1995A&A...299..657G} are present.
According to this scenario, B and D in Fig.~\ref{fig:warm_cold} could be the projection of the dust in the spiral arms that we see along the line of sight in the edge-on orientation, while A and E are the accumulation of H\textsc{ii} regions dominating the line of sight, harboring warm dust emitting at MIPS~-~24~$\mu$m (hardly visible at FIR and mm wavelengths; see Fig.~\ref{fig:sketch}). This picture agrees well with the decomposition reported by \cite{2012A&A...543A..74X} for M33. These differences can be better visualized in the bottom panel of Fig.~\ref{fig:warm_cold}, where the profiles along the major axis of the galaxy are overplotted, normalized at their values at 6 kpc (this region is free of any obvious enhanced emission in all bands). This plot clearly shows that region A stands out prominently at 24 and 70~$\mu$m, but is not significant at longer wavelengths (similar to region E, but much fainter at all wavelengths). Regions B and D are visible at all wavelengths (dimmer at mm wavelengths), while the central region, C, is prominent at all wavelengths, but is especially bright at 70~$\mu$m. Finally, region F at the extremes of the disk (11~kpc) clearly shows the relative dominance of very cold dust in this area. The NIKA2~-~1.15~mm emission is stronger than at other wavelengths, resulting in the emission excess relative to the fitted \texttt{HerBIE} model that we discussed in Sect.~\ref{subsec:resolvedSED} (see also Fig.~\ref{fig:mm_residuals}).

\subsection{Distribution of small and large dust grains}\label{sec:dust_grains}

The properties of the dust grains in \texttt{HerBIE} are described using the \texttt{THEMIS} model. \texttt{THEMIS} is based on laboratory data accounting for the aromatic and aliphatic MIR features with a single population of small partially hydrogenated amorphous carbons, noted a-C(:H). Although largely dehydrogenated, small a-C(:H) are very similar to polycyclic aromatic hydrocarbons (PAHs). The other main component of \texttt{THEMIS} is a population of large a-C(:H)-coated amorphous silicates, with Fe and FeS nano-inclusions. The distribution of aromatic-feature-emitting grains is parameterized in such a way as to distinguish between very small a-C(:H) (VSAC; smaller than 7~\AA) and small a-C(:H) 
(SAC; radius between 7~\AA~and 15~\AA) and between medium and large a-C(:H) 
(MLAC; with a radius larger than 15~\AA). 
Hereafter, the population of VSAC and SAC grains are referred to as small grains, and MLAC as large grains.
The free parameter that controls the mass fraction of small grains is $q_{AF} = q_{VSAC}+q_{SAC}$ , which, multiplied by the total dust mass, provides the dust mass of small grains. The mass of the small dust grains emitting in the NIR/MIR wavelengths is largely constrained by the \textit{Spitzer} and WISE observations, while the \textit{Herschel} and IRAM~-~NIKA2 measurements are necessary to constrain the emission from large grains that constitute the bulk of the dust mass \citep[see the emitting spectral regions of each component in Fig.~1 of][]{2021A&A...649A..18G}. In the case of NGC~891, the dust mass of small grains is calculated to be M$_{small~ grains}=3.32\times10^6$ M$_{\odot}$ (see Table~\ref{tab:global_sed}), which accounts for 9.5\% of the total dust mass.

%%%%%%%%%%%%%%%%%%%%%%%%%%%% FIGURE 8
\begin{figure}[t]
    \centering
    \captionsetup{labelfont=bf}
    \includegraphics[width=0.49\textwidth]{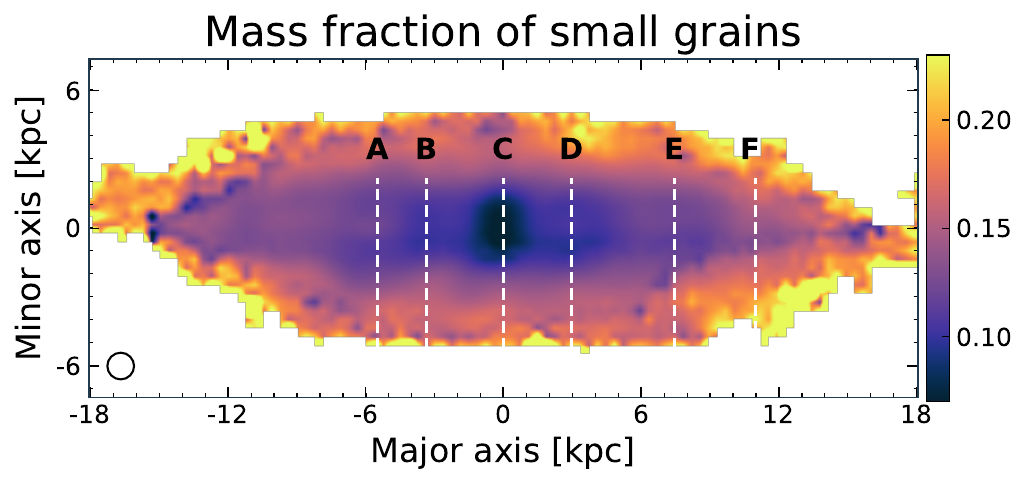}
    \caption{Mass fraction of the small dust grains (VSAC and SAC) over the total dust mass for NGC~891. The vertical dotted lines indicate the positions of interest introduced in Fig.~\ref{fig:seds}.}
    \label{fig:powerU}
\end{figure}

The map of the mass fraction of small grains is presented in Fig.~\ref{fig:powerU}. 
The mass fraction of small grains in most of the disk plane of the galaxy is $\sim10$\% (blue), with lower values found in the region of the bulge (region C) of the galaxy ($\sim6$\%), while regions of the disk with enhanced emission (regions A, B, D, and E) show increased abundances of small grains with mass fractions reaching up to $\sim15$\%.
At large distances above and below the plane of the disk ($>2$~kpc), the mass fraction of small dust grains increases and reaches up to $\sim20$\%. For comparison, the mass fraction of small a-C(:H) in the solar neighborhood from the \texttt{THEMIS} model is 17\% \citep{2021A&A...649A..18G}.

\begin{figure}[t]
    \centering
    \captionsetup{labelfont=bf}
    \includegraphics[width=0.49\textwidth]{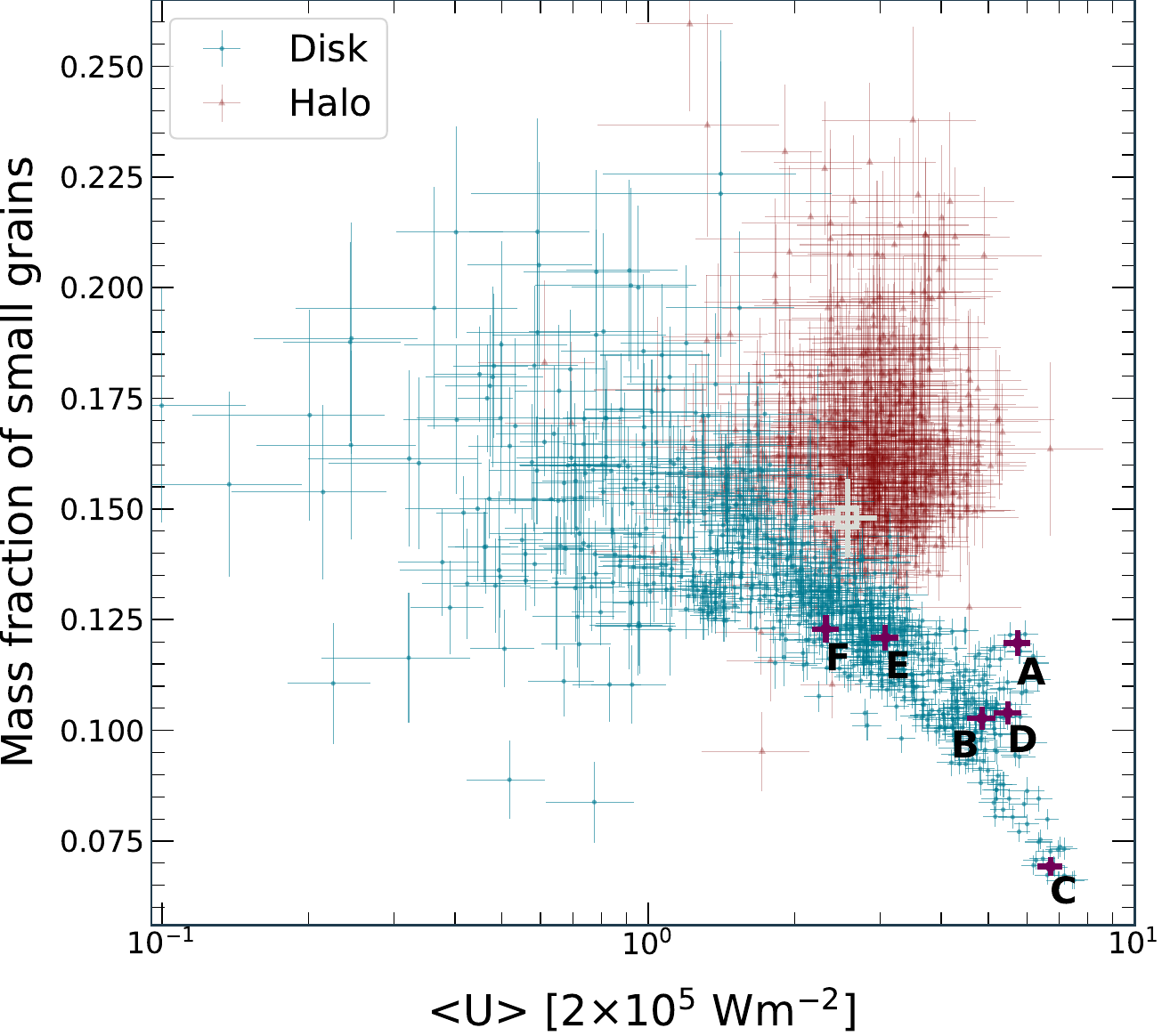}
    \caption{Mass fraction of the small dust grains (VSAC and SAC) as a function of the mean insterstellar radiation field <U>. 
    The cyan and red points are measurements for the individual pixels (along with their uncertainties) for the disk and the halo, respectively (with a crude separation at |z|~=~2~kpc). The purple symbols are median values and uncertainties within circular apertures with a radius of 0.8~kpc centered at the regions from A to F. The gray square in this plot shows the median value (and the respective uncertainty) of the dust emission feature located at $\sim4$~kpc above the center of the galaxy (see Fig.~\ref{fig:2mm}) within an elliptical aperture encompassing the feature.}
    \label{fig:correlation}
\end{figure}

Fig.~\ref{fig:powerU} indicates that the small dust grain abundance is low where the interstellar radiation field (ISRF) is intense and vice versa. In order to better quantify this anticorrelation, we plot in Fig.~\ref{fig:correlation} the mass fraction of small grains with the mean ISRF (<U>) from the \texttt{HerBIE} fit for every pixel. 
We focus on the A-F disk regions. Region A seems to be a particular outlier with respect to the rest of the disk regions: the mass fraction of the small dust grains is higher than that of the average ISRF. This agrees with the previous finding in Sect.~\ref{sec:dust_temp}, according to which this region is extremely bright at MIR wavelengths (indicating a large abundance of small grains) with very low emission of cold dust at FIR/submm emission. The enhancement of the fraction of the small grains at  high ISRF, such as in region A, might thus originate from the fact that some of these grains are shielded in the molecular cocoon of this giant HII region. In this plot, the dust feature at $\sim4$ kpc above the center of the galaxy (see Fig.~\ref{fig:2mm}) is in the locus between the disk and the halo regions (gray square and error bars), indicating that although it is located in the galactic halo, its properties are more similar to those of the disk. In this area, however, both the the mass fraction of small grains and the mean ISRF (<U>) have large uncertainties, which are also reflected in the relatively large error bars of this point compared to those calculated for regions A to F.

%%%%%%%%%%%%%%%%%%%%%%%%%%%% FIGURE 9

Fig.~\ref{fig:correlation} clearly shows that although the abundance of small dust grains is anticorrelated with <U>, two distinct populations (the disk and the halo) occupy two different loci in this plot. Small a-C(:H) could be destroyed within strong radiation fields. A negative correlation between the small grain abundance and <U> has already been reported \citep[e.g.,][]{2007ApJ...663..866D, 2008ApJ...672..214G, 2013MNRAS.431.2006K, 2015A&A...582A.121R, 2021A&A...649A..18G}. Therefore, a different processing mechanism may be active in the halo of the galaxy. \cite{2012A&A...541A..10Y} detected the 3.3~$\mu$m feature in the halo of M82, indicating the presence of small PAHs. It is possible that we obtain an excess of small a-C(:H) due to the shattering of larger carbon grains by shocks. The disk population, shown as cyan crosses in Fig.~\ref{fig:correlation}, is spread over at least one order of magnitude in <U> and a mass fraction of small grains ranging from $\sim0.06$ to $\sim0.2,$ while the halo population, shown as red crosses in Fig.~\ref{fig:correlation}, is very localized in the parameter space, with <U> having values within a relatively small range and a mass fraction of small grains ranging between $\sim0.125$ and $\sim0.2$. 

\begin{figure}[t]
    \centering
    \captionsetup{labelfont=bf}
    \includegraphics[width=0.49\textwidth]{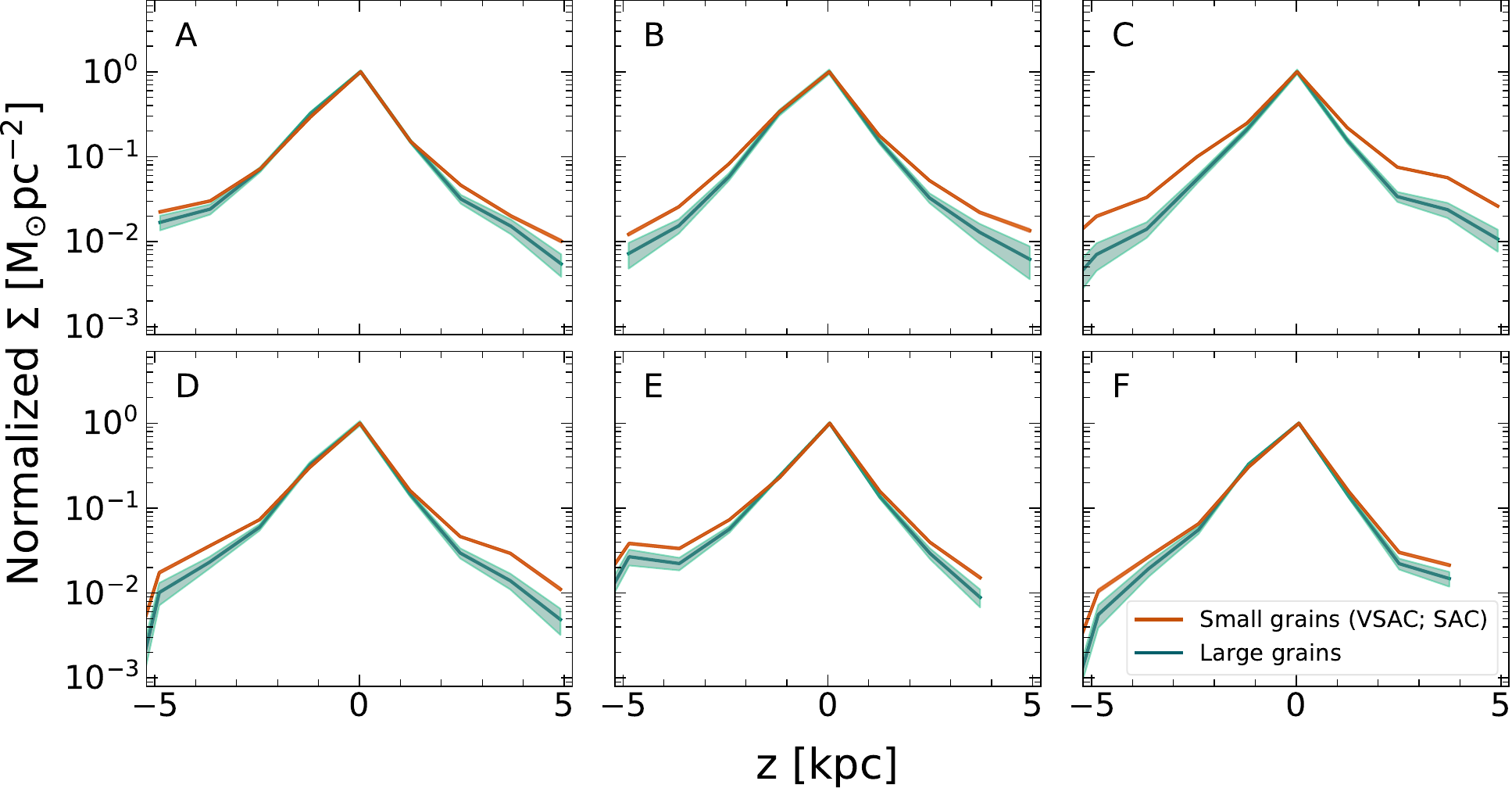}
    \caption{Vertical profiles of the mass surface density of small (VSAC and SAC; brown) and large grains (green) at different positions throughout the galactic disk. The profiles of the small dust grains are normalized so that they match the profiles of the large dust grains at z~=~0~kpc.}
    \label{fig:vertical_profiles}
\end{figure}

The increase in the abundance of small dust grains with galactic latitude 
(compared to large grains) is shown in Fig.~\ref{fig:vertical_profiles}, where the vertical profiles crossing the regions A, B, C, D, E, and F are plotted. The profiles of the small dust grains are scaled so that they match the profiles of the large dust grains at their central peaks. These profiles show that the small dust grains at $\sim|z|>2-2.5$~kpc (depending on the position) show a flatter distribution compared to the large grains. This difference is more prominent in the region C, where large grains decrease more steeply than small grains, which is evident already at $\sim1$~kpc.

In the disk, the negative correlation between the small grain fraction and the ISRF intensity (Fig.~\ref{fig:correlation}) probably results from the progressive destruction of these small grains by UV photons. This is indeed a well-known relation in the nearby Universe. Above the plane, the grains might be expelled from the galaxy via outflows. In these outflows, the shock waves could shatter the grains and thereby replenish the small grain reservoir. The contribution of this outflow to the total emission is negligible, but the edge-on orientation of NGC~891 allows us to see it precisely.

\cite{2014ApJ...785L..18S} reported a non-negligible amount of dust ($3-5$\% of the total dust mass) located at distances beyond 2~kpc from the midplane, in accordance with \cite{2016A&A...586A...8B}, who reported $2-3.3$\%. We find a higher mass fraction of small dust grains (8\%) for |z|~>~2~kpc. 
Furthermore, \cite{2016A&A...586A...8B} reported that the relative abundance of small grains with respect to the large grains (m$_{SG}$/m$_{LG}$) varied from 0.06 in the disk to 0.03 at a vertical distance of z~=~2.5~kpc. With our analysis, we find higher values of m$_{SG}$/m$_{LG}$ ranging from $\sim$ 0.07 to 0.15 in the disk and from $\sim$ 0.1 to 0.2 in the halo (at distances larger than 2~kpc from the midplane). Our result that the ratio of m$_{SG}$/m$_{LG}$ increases with distance from the midplane differs from previous studies. It is probable that the different modeling techniques used or the different wavelength coverage \citep[e.g., in][the large grain emission is constrained by wavelengths only up to 250~$\mu$m]{2016A&A...586A...8B} can explain these discrepancies.

%%%%%%%%%%%%%%%%%%%%%%%%%%%% FIGURE 10

\section{Conclusions} \label{sec:conclu}
We explored the complex mm wavelength range of the SED of the edge-on nearby galaxy NGC~891. We presented new observations of the galaxy at 1.15 and 2~mm obtained with the NIKA2 camera on the IRAM~30m telescope. Making use of these unique mm data combined with a set of multi-wavelength ancillary data and the \texttt{HerBIE} SED fitting code, we conclude the following:

-- By comparing the modeled maps at 1.15 and 2~mm, we find that there is significant evidence of submm/mm dust emission in excess compared to the dust model in the outermost regions of the galaxy (between 10 and 12~kpc). The NIKA2 fluxes are higher by up to 25\% than the model-predicted emission.  

-- Decomposing the emission at mm wavelengths into dust, free-free, and synchrotron, we find different morphologies of each component in the disk and in the halo of the galaxy. Specifically, at 2~mm, the disk emission shows prominent dust and synchrotron emission in the bulge region without obvious free-free emission. Regions of enhanced cold dust emission away from the center of the galaxy do not seem to correlate well with free-free and synchrotron emission features. Bright free-free emission features in the disk show deficits in the synchrotron map. On the other hand, the dust halo of the galaxy maintains an elliptical shape similar to that of the disk, even at high galactic latitudes, while the synchrotron and free-free maps show a peanut-like shaped halo.

-- The emission at mm/cm wavelengths was decomposed in detail, and the relative contributions of the dust, the free-free, and the synchrotron emissions were presented for regions in the disk and in the halo of the galaxy. At 1~mm, the emission comes from the dust (with negligible emission coming from free-free emission), and the free-free emission begins to increase at 2~mm at levels of $\sim5-15\%$ depending on their location in the disk or halo. At 5~mm, the radio emission becomes the primary component, and the free-free emission dominates in the disk, where it reaches emission levels up to $\sim70\%,$ while both free-free and synchrotron emission dominate in the halo. Dust at 5~mm is still a significant emission source with a surprisingly larger contribution in the halo rather than in the disk ($\sim30-40$\% in halo compared to $\sim20-30$\% in the disk). At 2~cm, only radio emission with the free-free emission dominates the disk and the synchrotron emission the halo, while at 20~cm most of the emission in the disk and in the halo of the galaxy is synchrotron emission.

-- Comparing the dust morphology seen at warm and cold dust tracers, we detected enhanced emission features in the galactic disk seen at all wavelengths (e.g., the central bulge region), only at MIR wavelengths, and only at FIR/mm wavelengths. We explained this by a simple scenario in which cold regions are the product of the projection in the edge-on geometry of the cold dust situated along the spiral arms of the galaxy, while warm dust emission arises from dust in large compact H\textsc{ii} regions.

-- Taking advantage of the analysis performed with the \texttt{HerBIE} code, we decomposed the dust emission into two components, one component originating from small dust grains (smaller than 15~\AA), and the other originating from the larger grains. We concluded that the mass fraction of the small grains accounts for $\sim9.5$\% of the total dust mass. In the disk of the galaxy, the mass fraction of small dust grains varies from $\sim6$\% in the center of the galaxy to $\sim10$\% in other regions in the disk, reaching up to $\sim15$\% of enhanced dust emission.
At large distances above and below the disk (>~2~kpc), the mass fraction of the small dust grains increases, reaching up to 20\%. The distribution of small grains from the disk into the halo is flatter than the distribution of the large grains, indicating the increase in the relative abundance of this dust population at high galactic latitudes. We speculate that the grains are expelled from the galaxy via outflows in which the shock waves shatter the grains and thereby replenish the small grain reservoir.

The NIKA2 observations at 1.15 and 2~mm of the nearby galaxy NGC~891 have proven to be very important in constraining the SED of the galaxy at mm wavelengths and to investigate the morphology of the cold dust with respect to warmer dust as well as the radio emission. Similar studies, exploiting the NIKA2 observations of nearby galaxies observed within the framework of the IMEGIN program, are expected to further advance our understanding of the ISM properties and of the thermal/radio emission mechanisms that are taking place in galaxies.

\begin{acknowledgements}
We would like to thank the referee, Santiago Garcia-Burillo, for the useful comments and suggestions, which helped improving the quality of the manuscript. The research work was supported by the Hellenic Foundation for Research and Innovation (HFRI) under the 3rd Call for HFRI PhD Fellowships (Fellowship Number: 5357). 
We would like to thank the IRAM staff for their support during the campaigns. The NIKA2 dilution cryostat has been designed and built at the Institut N\'eel. In particular, we acknowledge the crucial contribution of the Cryogenics Group, and in particular Gregory Garde, Henri Rodenas, Jean Paul Leggeri, Philippe Camus. 
The NIKA2 data were processed using the Pointing and Imaging In Continuum (PIIC) software, developed by Robert Zylka at the Institut de Radioastronomie Millim\'etrique (IRAM) and distributed by IRAM via the GILDAS pages. PIIC is the extension of the MOPSIC data reduction software to the case of NIKA2 data. 
This work has been partially funded by the Foundation Nanoscience Grenoble and the LabEx FOCUS ANR-11-LABX-0013. This work is supported by the French National Research Agency under the contracts "MKIDS", "NIKA" and ANR-15-CE31-0017 and in the framework of the "Investissements d’avenir" program (ANR-15-IDEX-02). 
This work has benefited from the support of the European Research Council Advanced Grant ORISTARS under the European Union's Seventh Framework Programme (Grant Agreement no. 291294). 
F.R. acknowledges financial supports provided by NASA through SAO Award Number SV2-82023 issued by the Chandra X-Ray Observatory Center, which is operated by the Smithsonian Astrophysical Observatory for and on behalf of NASA under contract NAS8-03060. 
This work was supported by the Programme National "Physique et Chimie du Milieu Interstellaire" (PCMI) of CNRS/INSU with INC/INP and the Programme National Cosmology et Galaxies (PNCG) of CNRS/INSU with INP and IN2P3, co-funded by CEA and CNES. 
M.B., A.N., and S.C.M. acknowledge support from the Flemish Fund for Scientific Research (FWO-Vlaanderen, research project G0C4723N).
\end{acknowledgements}

\bibliographystyle{aa_url}
\bibliography{References}

%%%%%%%%%%%%%%%%%%%%%%%%%%%%%%%%%%%%%%%%%%%%%%%%
%%%%%%%%%%%%%%%%% APPENDIX %%%%%%%%%%%%%%%%%%%%%
%%%%%%%%%%%%%%%%%%%%%%%%%%%%%%%%%%%%%%%%%%%%%%%%
\begin{appendix}
\onecolumn
\section{Model-derived parameters for the regions of interest}\label{sec:appendA}
Table~\ref{tab:spatial_sed} lists the parameters derived with our analysis in regions A to F (see Fig~\ref{fig:seds}). The values are averages within circular apertures with a radius of 1~kpc centered at these regions. The typical SEDs in these regions of interest are shown in Fig.~\ref{fig:seds}.
    \begin{table}[h]
    \centering
    \captionsetup{labelfont=bf, labelformat=simple} 
    \begin{tabular}{lcccccc}
    \hline\hline
    Parameters   & A & B & C & D & E & F \\ \hline
    $\Sigma_{dust}$~[M$_{\odot}$~pc$^{-2}$]        & 0.828~$\pm$~0.048  & 1.068~$\pm$~0.061  & 1.154~$\pm$~0.063  & 0.948~$\pm$~0.059  & 0.678~$\pm$~0.040  & 0.569~$\pm$~0.034 \\
    $\Sigma_{small~grains}$~[M$_{\odot}$~pc$^{-2}$] & 0.098~$\pm$~0.006  & 0.111~$\pm$~0.007  & 0.086~$\pm$~0.005  & 0.097~$\pm$~0.007  & 0.081~$\pm$~0.005  & 0.070~$\pm$~0.005 \\
    $\Sigma_{large~grains}$~[M$_{\odot}$~pc$^{-2}$] & 0.730~$\pm$~0.042  & 0.958~$\pm$~0.055  & 1.068~$\pm$~0.058  & 0.851~$\pm$~0.053  & 0.597~$\pm$~0.035  & 0.499~$\pm$~0.030  \\
    $\Sigma_{gas}$/$\Sigma_{dust}$   &  239~$\pm$~27  & 203~$\pm$~23  & 170~$\pm$~20  & 237~$\pm$~28  & 196~$\pm$~23  & 184~$\pm$~21   \\
    T$_{dust}$~[K]        & 24.50~$\pm$~0.27  & 24.02~$\pm$~0.26  & 25.17~$\pm$~0.27  & 24.41~$\pm$~0.28  & 22.28~$\pm$~0.25  & 21.10~$\pm$~0.23   \\
    L$_{star}$~[L$_{\odot}$~pc$^{-2}$]        &  804~$\pm$~24  & 1260~$\pm$~29  & 3075~$\pm$~55  & 1229~$\pm$~29  & 632~$\pm$~15  & 422~$\pm$~10  \\
    L$_{dust}$~[L$_{\odot}$~pc$^{-2}$]        & 902~$\pm$~19  & 1018~$\pm$~21  & 1400~$\pm$~32  & 989~$\pm$~21  & 427~$\pm$~9  & 268~$\pm$~5 \\
    <U>~[$2.2\times10^5$ Wm$^{-2}$]          & 5.46~$\pm$~0.34  & 4.84~$\pm$~0.31  & 7.01~$\pm$~0.45  & 5.31~$\pm$~0.36  & 3.14~$\pm$~0.21  & 2.32~$\pm$~0.15 \\
    q$_{AF}$          &  0.117~$\pm$~0.003  & 0.104~$\pm$~0.003  & 0.076~$\pm$~0.002  & 0.102~$\pm$~0.003  & 0.119~$\pm$~0.003  & 0.123~$\pm$~0.003 \\
    f$_{ion}$         & 0.497~$\pm$~0.016  & 0.517~$\pm$~0.018  & 0.573~$\pm$~0.026  & 0.568~$\pm$~0.017  & 0.546~$\pm$~0.017  & 0.519~$\pm$~0.016 \\
    $\beta$         & 2.481~$\pm$~0.027  & 2.483~$\pm$~0.017  & 2.487~$\pm$~0.012  & 2.484~$\pm$~0.018  & 2.482~$\pm$~0.018  & 2.481~$\pm$~0.015 \\
    $\alpha_s$           &  0.894~$\pm$~0.003  & 0.894~$\pm$~0.003  & 0.895~$\pm$~0.002  & 0.894~$\pm$~0.003  & 0.894~$\pm$~0.003  & 0.894~$\pm$~0.003 \\ 
\hline
    \end{tabular}
    \caption{Values of the model parameters in regions A to F (see  Fig.~\ref{fig:seds}).}
    \label{tab:spatial_sed}
    \end{table}
\end{appendix}

\end{document}